\documentclass[aps, superscriptaddress, longbibliography,pre,10pt]{revtex4-1}
\usepackage[font=small,labelfont=bf]{caption}
\usepackage[large]{subfigure }
\usepackage{amsmath,graphicx,color,epsfig,latexsym,bm,ulem,float,tikz,eucal, mathpazo,times, braket,comment}
\usepackage[colorlinks=true, linkcolor = blue, urlcolor  = blue, citecolor = blue, anchorcolor = red]{hyperref}   

\usepackage{lipsum}
\makeatletter
\newcommand{\shorteq}{%
	\settowidth{\@tempdima}{=}
	\resizebox{\@tempdima}{\height}{=}%
}
\makeatletter
\newcommand*\bigcdot{\mathpalette\bigcdot@{.5}}
\newcommand*\bigcdot@[2]{\mathbin{\vcenter{\hbox{\scalebox{#2}{$\m@th#1\bullet$}}}}}
\makeatother
\makeatother
\usetikzlibrary{positioning}

\begin{document}
	
	\title{Current circulation near additional energy degeneracy points in quadratic Fermionic networks}
	\author{Vipul Upadhyay} \email{vipuupadhyay4@gmail.com} 
 \thanks{Currently affiliated with the Department of Chemistry, Institute of Nanotechnology and Advanced Materials, Center for Quantum Entanglement Science and Technology, Bar-Ilan University, Ramat-Gan, 52900, Israel.}
	\affiliation{Department of Physics, Indian Institute of Technology Delhi, Hauz Khas 110 016, New Delhi, INDIA}

	\author{Rahul Marathe}\email{maratherahul@physics.iitd.ac.in} \affiliation{Department of Physics, Indian Institute of Technology Delhi, Hauz Khas 110 016, New Delhi, INDIA}
	
	\begin{abstract}{We study heat and particle current circulation (CC) in quadratic Fermionic systems analysed using a general dissipative Lindbladian master equation. It was observed in an earlier study \cite{Our_circulation} that CC occurs near the additional energy degeneracy point (AEDP) in Fermionic systems which have some form of asymmetry. We find general analytical expression to support this observation for quadratic Fermionic networks.  We then apply these ideas to the  Su–Schrieffer–Heeger (SSH) model with periodic boundary conditions and a tight binding model with unequal  hopping  strengths in the upper and lower branches. In both these cases, we find the specific conditions required for observing CC and study the behavior of these currents with various system parameters.  We find that having unequal number of Fermionic sites in the upper and lower branches is enough for generating CC in the SSH model. However, this asymmetry is not adequate for the tight-binding model and we require unequal hopping strengths in the upper and lower branches to induce CC in this model. {We also compare our results with the exact results obtained via the Non-Equilibrium Green Function (NEGF) formalism, and observe that the relationship between AEDP and CC also holds for the exact results}. Finally, we observe that for certain system parameters, the onset point of particle and heat CC are not the same. Based on all these observations, we describe how carefully examining the energy spectrum of the system gives a great deal of information about the  possibility and behavior of CC in Fermionic systems with asymmetries.}
	\end{abstract}
	
	\maketitle
	\section{Introduction}
	Current circulation (CC) or Current magnification (CM) is an exotic transport phenomenon observed in multi-branched systems kept between two heat baths with a temperature gradient. For the simplest system with two current-carrying branches between two thermal baths, it occurs when the current in one of the branches becomes larger than the total current entering the system \cite{main_reference,Our_circulation}.
	As such, the other branch is forced to have a current flow in the opposite direction to the thermal gradient and the current circulates either clockwise or anticlockwise in the branches.
	\\ CC is a ubiquitous wave phenomenon and has been theoretically observed in a wide variety of physical systems like quantum spins \cite{Our_circulation,main_reference,Unequal_CC_Yue}, mesoscopic rings \cite{CC_jayannavar,CC_jayavannar_2,current_magnification5}, molecular wires \cite{CC_mplecular_wires_abraham_nitzan}, classical harmonic and anharmonic oscillators \cite{Germometric_CC_1_PhysRevE.99.022131,CC_classical_rahul_sir,circulation_zabey_coupled_Oscillator}, Fermionic, Bosonic and other realisations of quantum dots \cite{triangular_dot_EPL,Geometric_2_PhysRevA.102.023704,Geometry_3_PhysRevE.105.064111,current_magnification4}. The usual justification for observing CC in quantum systems has been accounted to a phenomenon called Fano resonance \cite{Fano_ref,CC_jayannavar}  which makes contributions from some current conducting channels significantly larger near specific  parameter regimes. In recent studies \cite{Germometric_CC_1_PhysRevE.99.022131,Geometric_2_PhysRevA.102.023704,Geometry_3_PhysRevE.105.064111,Our_circulation,Geometric_gassab2023geometrical}, the role of geometrical asymmetries in observing anomalous transport like CC has been highlighted.  Interestingly, it has also been observed that in quantum systems with some asymmetry analysed using the master equation approach, the onset of CC is usually near the additional energy degeneracy points (AEDPs) in the system \cite{Our_circulation, main_reference}. These points refer to the intersection points of two otherwise distinct energy levels in the system. Motivated by these findings,  we explore the reason behind the existence of  CC near the AEDP in  quadratic Fermionic systems in this manuscript. Using a local Lindblad master equation \cite{Breuer_PhysRevE.76.031115} formulation for analysing  the system dynamics, we provide a  general analytical framework demonstrating the possibility of CC in these systems. We then study the application of this idea in the SSH model \cite{ssh_original_paper,short_course_topo,sshreview} with periodic boundary conditions and the tight-binding model with asymmetric hopping strength in the upper and lower branches. We find that CC for these models does occur near AEDP but some form of asymmetry in the system is required. We then employ different geometrical asymmetries in these models and highlight the  specific conditions required for observing CC. Interestingly, we find that for certain systems, the onset point of particle CC and heat CC may not be the same and we may have scenarios where the heat current circulates while the particle current runs parallel in the branches. We explore the conditions required for observing this effect. The experiments related to the models studied in this manuscript can be performed in setups like cold atoms\cite{exp_cold_atoms,exp_cold_atoms1}, trapped ions \cite{exp_trapped_ion,exp_trapped_ion2}, engineered atomic lattices \cite{exp_EAL_1,exp5}, quantum simulators like NMR \cite{NMR2,Nature_experiment_NMR} and quantum dots  \cite{exp_quantum_dots,exp_quantum_dot_2}. Although the transport properties of open non interacting Fermionic chains have been intensely studied analytically \cite{Review_dario_RevModPhys.94.045006,Refree_1PhysRevE.89.022102,Ref2PhysRevA.95.052107,Ref3PhysRevA.98.052126,Ref4Prosen_2010,Ref5Prosen_2010,Ref610.21468/SciPostPhys.14.5.112}, the importance of our study lies in its focus on the relationship between AEDP and current circulation in periodic chains. Additionally, we utilize a perturbative approach to derive the analytical expression for the currents. Similar methods have been previously employed, by calculating the density matrix order by order in terms of the system bath interaction strength \cite{Juzar_PhysRevB.85.195452,Juzar_10.1063/1.4718706},  or more recently, by weakening the bonds connected to the bath and using this as the perturbation parameter \cite{Lyapunov_PhysRevE.94.032139,Lyapunov_PhysRevB.107.035113}.  In our case, we take the system-bath coupling strength as the perturbation parameter. This means that we are looking at those setups where the system and the bath are weakly coupled. 
	\par The manuscript is organised as follows. We introduce the non-interacting Fermionic Hamiltonian  and discuss the general dissipative Lindblad master equation (GLME), and the correlation matrix master equation in section \ref{General_System_Hamiltonian}. We then provide a perturbative framework for solving the correlation matrix master equation in section \ref{solving_the_local_QMES}, and apply it to provide a general criterion under which CC may occur in  our system.  Using the local Lindblad Master Equation (LLME) as an  example, the application of this idea to the SSH model with unequal number of Fermionic sites in upper and lower branches is studied in section \ref{SSH_Model}. A comparison between the approximate results of LLME and the exact results obatined via the Non-Equilibrium Green Function (NEGF) formalism is given in section \ref{NEGF_results}.
This is followed by the application of the idea to the tight-binding model with asymmetric hopping strengths in upper and lower branch in section \ref{Asymm_Model}. This is followed by a brief discussion on the effects of adding interactions to the system in section \ref{Interaction_section}. This is done by adding a Hubbard model \cite{Hubbard_review_LIEB20031} like interaction term to the earlier SSH model.  Finally, based on the observation from the results in these two models we draw conclusions in section \ref{Conclusion}.
		
		\section{System Hamiltonian and Master Equation} \label{General_System_Hamiltonian}
		We consider quadratic Fermionic  Hamiltonians  of the following form, 
		\begin{align}\label{model_hamiltonian}
			\hat{H}_S&=\sum_{i\ne j=1}^{N} v_{i j} \hat{c}_{i}^\dagger \hat{c}_j+\Omega \sum_{n=1}^{N}\hat{c}_n^{\dagger}\hat{c}_n
		\end{align}
		where $\hat{c}_i^\dagger (\hat{c}_i)$ are the Fermionic creation and destruction operators at site $i$, $v_{i j}$ is the hopping strength from site $i$ to site $j$ and $v_{i,j}=v_{j,i}$, $\Omega$ is the chemical potential at  all sites and we have periodic boundary conditions, so, $\hat{c}_{N+1}=\hat{c}_1$.  This form of the Hamiltonian preserves the total number of Fermions and only allows Fermions to hop from an occupied site to an unoccupied site at a rate proportional to the hopping  strength. As such, direct interaction between Fermions is not considered in such Hamiltonians \cite{sshreview}. Such Hamiltonians can be written in terms of a matrix $\hat{M}$ as,
		\begin{align} \label{M_define}
				\hat{H}_S= \sum_{i,j} M_{i,j} \hat{c}_i^\dagger \hat{c}_j
			\end{align}
			where, $\hat{M}_{i,j}=v_{i,j}+\Omega$ is the $N\times N$ Hamiltonian matrix. The system is connected at two nodal sites to two  different Fermionic baths at different temperatures, having the following Hamiltonian,
		\begin{align}
			\hat{H}_B=\sum_n \omega^L_n \hat{b}_n^{L \dagger}\hat{b}^L_n+\sum_{n}\omega^R_n \hat{b}_n^{R \dagger}\hat{b}^R_n
		\end{align}	
		where $\hat{b}_n^{i \dagger} (\hat{b}_n^{i})$ is the creation (destruction) operator of the  $n^{th}$ mode of the $i^{th}$ bath. The system bath coupling has the following form,
		\begin{align}
			\hat{H}_{SB}=&\hat{c}_1^{\dagger}\otimes  \sum_k g^L_k  \hat{b}^L_k + \hat{c}_1\otimes  \sum_k g^{L*}_k  \hat{b}_k^{L\dagger} +\hat{c}_{N_U+2}^{\dagger}\otimes  \sum_k g^R_k  \hat{b}^R_k + \hat{c}_{N_U+2}\otimes  \sum_k g^{R*}_k  \hat{b}_k^{R\dagger} 
		\end{align}
		where $g^L_k ( g^R_k)$ is the system bath interaction strength for left (right) terms and the bath is connected at the sites numbered $1$ and $N_U+2$.
		\subsection{Master equation}
		
			We consider a general Lindblad master equation (GLME) for analysing our system \cite{breuer2002} . It has the following form for the Hamiltonians given in the above section,
			\begin{align}\label{local_master_equatio}
				\frac{d \hat{\rho}}{dt}&=-i[\hat{H}_S,\hat{\rho}]+ \sum_i \gamma_i \mathcal{D}(\hat{c}_i)[\hat{\rho}]+\sum_i\Gamma_i\mathcal{D}(\hat{c}_i^{\dagger})[\hat{\rho}]
			\end{align}
			where, we have assumed general on site dissipators, $\hat{c}_i (\hat{c}^\dagger_i)$. The respective rates for these dissipators are given as, $\gamma_i (\Gamma_i)$. These rates may depend upon both the Hamiltonian and the bath parameters, and their exact form depends on how the Lindblad master equation is derived \cite{breuer2002,Breuer_PhysRevE.76.031115,Upadhyay_2024}. The Lindbladian dissipators have the following form,
			\begin{align}
				\mathcal{D}(\hat{A})[{{\rho}}]=\hat{A} {{\rho}} \hat{A}^{\dagger}- \frac{1}{2}\big{(}{{\rho}} \hat{A}^{\dagger}\hat{A} + \hat{A}^{\dagger} \hat{A} {{\rho}}\big{)}
			\end{align}
			\subsection{Correlation Matrix Master equation}
			We now try to find the master equation for the Correlation matrix $\hat{C}$ defined below \cite{SPDM},
			\begin{align}
				\hat{C}_{i,j} \equiv Tr[\hat{c}_i^\dagger\hat{c}_j {{\rho}}]
			\end{align}
			Using eq. \eqref{local_master_equatio},	the equation governing the matrix element of the correlation matrix becomes:
			\begin{align}
				\frac{d Tr[\hat{c}_i^\dagger \hat{c}_j \hat{\rho}]}{dt}=&Tr\Big{[}\hat{c}_i^\dagger \hat{c}_j (-i[\hat{H}_S,\hat{\rho}]+\sum_n \gamma_n \mathcal{D}(\hat{c}_n)[\hat{\rho}]+ \sum_n  \Gamma_n\mathcal{D}(\hat{c}_n^{\dagger})[\hat{\rho}])\Big{]}
			\end{align}
			Expanding $\hat{H}_S$ and the Lindbladian dissipators in terms of Fermionic operators and using the anticommutation property of the Fermionic operators,
			it can be shown that \cite{Lyapunov_PhysRevB.107.035113} (see Appendix \ref{Sec:AppendixA}),
			\begin{align}\label{Correlation_Matrix_dynamic}
				\dot{\hat{C}}=i [\hat{M}^T,\hat{C}]+\hat{G}-\{\frac{(\hat{G}+\hat{R})}{2},\hat{C}\}
			\end{align}
			where $\hat{M}^T$ is the transpose of the Hamiltonian matrix $M$. Since the Hamiltonian is symmetric, $\hat{M}^T=\hat{M}$. The matrices $\hat{G}$ and $\hat{R}$ are given as,
			\begin{align}
				\hat{G}_{i,j}&=\delta_{i,j}\Gamma_j,
				&&\hat{R}_{i,j}=	\delta_{i,j}\gamma_j.
			\end{align}
			Now, if we substitute $	\hat{A}=\frac{\hat{G}+\hat{R}}{2}-i \hat{M}^T$, we get the following Lyapunov equation for the steady state correlation matrix \cite{Lyapunov_PhysRevE.94.032139},
			\begin{align} \label{Lyapunov}
				\hat{A}\hat{C}+\hat{C}\hat{A}^\dagger=\hat{G}
			\end{align}
			where the operators,
			\begin{align}
				&	\hat{A}=\frac{\hat{G}+\hat{R}}{2}-i \hat{M}^T,&&
				\hat{A}^\dagger=\frac{\hat{G}+\hat{R}}{2}+i \hat{M}		
			\end{align}	
			Looking at the form of the $\hat{A}$ matrix, we see that it contains information about both the system Hamiltonian through the matrix $\hat{M}$ and the system's interaction to left and right bath through matrices $\hat{G}$ and $\hat{R}$. \textbf{Note:} Though the above correlation matrix master equation is written only for quadratic Fermionic systems, but replacing the Fermionic operators with the Bosonic operators gives  the same equation for Bosonic correlation matrix (see Appendix \ref{Appendix_B}). Hence, we expect the conclusions for quadratic Fermionic systems to also hold true for quadratic Bosonic systems.
	
	\section{Solving The Master Equation for diagonalisable $\hat{A}$} \label{solving_the_local_QMES}
	In this study, we assume that the Hamiltonian and system bath interaction are chosen such that  the matrices $\hat{A}^\dagger$  are  always diagonalisable, so we can always perform the following diagonalisation transformation \cite{Arfken}
	\begin{align}
		&	\hat{D}=\hat{P}^{-1}\hat{A}^\dagger\hat{P}
	\end{align}
	where, $\hat{D}$ is a diagonal matrix having the eigenvalues of $\hat{A}^\dagger$ at its diagonal and $\hat{P}$ is the matrix doing the above transformation. Using this, it can be shown that the eq. \eqref{Lyapunov} becomes,
	\begin{align}
		\hat{D}^\dagger \tilde{C}+\tilde{C} \hat{D}=\tilde{G}
	\end{align}
	where $\tilde{C}$ is the following transformation of the $\hat{C}$ matrix,
	\begin{align}\label{c_basis_conversion}
		&\tilde{C}=\hat{P}^\dagger \hat{C} \hat{P}
	\end{align}
	The same transformation holds true for $\tilde{G}$. From the above equation, we can formally write the solutions for matrix elements of $\tilde{C}$ as,
	\begin{align} \label{c_tilda_exact}
		\tilde{C}_{i,j}=\frac{\tilde{G}_{i,j}}{\hat{D}^\dagger_{i,i}+\hat{D}_{j,j}}
	\end{align}	
	Reversing the transformation in eq. \eqref{c_basis_conversion}, we can find the correlation matrix in the particle-hole basis.	
	
	\subsection{Expression for Currents}
	Typically, the expression for local current flowing in a non-interacting system depends only on the Correlation matrix and is proportional to the following  matrix element  \cite{Prosen_2008_PhysRevLett.101.105701,Lyapunov_PhysRevB.107.035113,Lyapunov_PhysRevE.94.032139},
	\begin{align}
		&i (\hat{C}_{i,j}-\hat{C}_{j,i})=2 i Im (\hat{C}_{i,j})=2 i Im((\hat{P}^\dagger)^{-1} \tilde{C} \hat{P}^{-1})_{i,j}
	\end{align}
	Solving, using eq. \eqref{c_basis_conversion} and \eqref{c_tilda_exact} we get,
	\begin{align} \label{Cij_eq}
		i (\hat{C}_{i,j}-\hat{C}_{j,i})&=2 i Im\sum_{l,m}(\hat{P}^\dagger)^{-1}_{i,l}  \frac{\tilde{G}_{l,m}}{\hat{D}^\dagger_{l,l}+\hat{D}_{m,m}}\hat{P}^{-1}_{m,j}
	\end{align}
	where,	$\tilde{G}=\hat{P}^\dagger \hat{G} \hat{P}$. 
	
	\subsection{Perturbation theory}\label{circulation_possible}
	The above analysis tells us that if we can find the eigenvalues and eigenvectors of the $\hat{A} (\hat{A}^\dagger)$  matrix, we can find the local currents in the system. The rates present in the Martices $\hat{G},\hat{R}$ are dependent on the square of the system bath coupling strength, ($\kappa \propto |g^{L(R)}_K|^2$)  \cite{breuer2002,Breuer_PhysRevE.76.031115}. So, we can write the matrix $\hat{A}^\dagger$ in the following form,
	\begin{align} \label{hat_A_define}
		\hat{A}^\dagger=i \hat{M}+\frac{\hat{G}+\hat{R}}{2}=\hat{M}_0+\kappa \hat{M_1}
	\end{align}			
	where $\hat{M}_0=i \hat{M}, \hat{M}_1=\frac{\hat{G}+\hat{R}}{2 \kappa}$. Here we make the $\kappa$ dependence of matrix $\hat{A}^\dagger$ explicit.  Now, $\kappa$  is usually very small for the master equation we are considering, so we can  perform a perturbative analysis to find the eigenspecturm of $\hat{A}^\dagger$ in terms of the perturbation parameter $\kappa$,  
	\begin{align}
		\hat{A}^\dagger |\lambda_i \rangle=\hat{M}_0+\kappa \hat{M}_1|\lambda_i \rangle =D_{ii} | \lambda_i \rangle 
	\end{align}
	The matrix $\hat{M}_0 =i \hat{M}$, with $\hat{M}$ being Hamiltonian matrix of eq. \eqref{M_define}. As such, it is a skew-symmetric matrix with eigenvalues $`i'$ times the energy eigenvalues, and has the same eigenvectors  as $\hat{M}$. Considering this, we expand  $D_{ii}$ and $\lambda_i$ up to the first order in the power of $\kappa$.
	\begin{align}
		D_{ii}&=iE_i+\kappa E_i^{(1)}+\mathcal{O}(\kappa^2)\ldots \nonumber \\
		|	\lambda_{i}\rangle &=|\lambda^{(0)}_i\rangle+\kappa |\lambda_i^{(1)}\rangle+\mathcal{O}(\kappa^2)\ldots
	\end{align}
	where, $E_i$ is the energy corresponding to $i^{th}$ energy eigenvector of the system.  The expansion terms can be found by using the time-independent perturbation theory \cite{griffiths_schroeter_2018}.  If the levels $E_i$ are non-degenerate,
	\begin{align}
		E_{i}^{(1)}=\langle \lambda_i^{(0)} |\hat{M}_1| \lambda_i^{(0)}\rangle
	\end{align}
	and if the levels $E_a, E_b$ are degenerate, the first order corrections are,
	\begin{align}
		E_{a,b}^{(1)}=\pm \frac{1}{2}\Big{[}W_{aa}+W_{bb}\pm\sqrt{(W_{aa}-W_{bb})^2+4|W_{ab}|^2}\Big{]}
	\end{align}
	where $W_{ab}=\langle \lambda^{(0)}_a |\hat{M}_1| \lambda^{(0)}_b\rangle$. If the first order corrections are real, the equation \eqref{Cij_eq} can be approximately written as,
	\begin{align}
		&i (\hat{C}_{i,j}-\hat{C}_{j,i})\approx 2 i Im\sum_{l,m} \frac{ (\hat{P}^\dagger)^{-1}_{i,l} \tilde{G}_{l,m} \hat{P}^{-1}_{m,j}}{\kappa(E^{(1)}_l+E^{(1)}_m )+i (E_m-E_l)}
	\end{align}
	where, the denominator is correct upto the first order in the system-bath coupling paramter $\kappa$. Suppose we perform the following expansion,
	\begin{align}
		(\hat{P}^\dagger)^{-1}_{i,l} \tilde{G}_{l,m} \hat{P}^{-1}_{m,j}	=(\epsilon_R)_{l,m}^{i,j}+i (\epsilon_I)_{l,m}^{i,j}
	\end{align}
	where $(\epsilon_R)_{l,m}^{i,j}$ and $(\epsilon_I)_{l,m}^{i,j}$ are the real and imaginary parts of the complex number  $(\hat{P}^\dagger)^{-1}_{i,l} \tilde{G}_{l,m} \hat{P}^{-1}_{m,j}$, we have:	
	\begin{align}\label{current_contribution}
		&i (\hat{C}_{i,j}-\hat{C}_{j,i})\approx -2 \sum_{l,m} \frac{(\epsilon_R)_{l,m}^{i,j}(E_m-E_l)-(\epsilon_I)_{l,m}^{i,j}\kappa(E^{(1)}_l+E^{(1)}_m)}{\kappa^2(E^{(1)}_l+E^{(1)}_m)^2 + (E_m-E_l)^2}
	\end{align}
	Looking at the above expression for local current, we see that the contribution to it from the interaction between the energy levels $l,m$ is in the following functional form,
	\begin{align} \label{Fano resonance form}
		F(a,b,E^{(1)}_S, \Delta E)= \frac{a \Delta E- b E^{(1)}_S}{ \Delta E^2 +E^{(1)2}_S}
	\end{align}
	\begin{figure}[t]
		\centering 
		\subfigure []
		{\includegraphics[width=0.4\linewidth,height=0.3\linewidth]{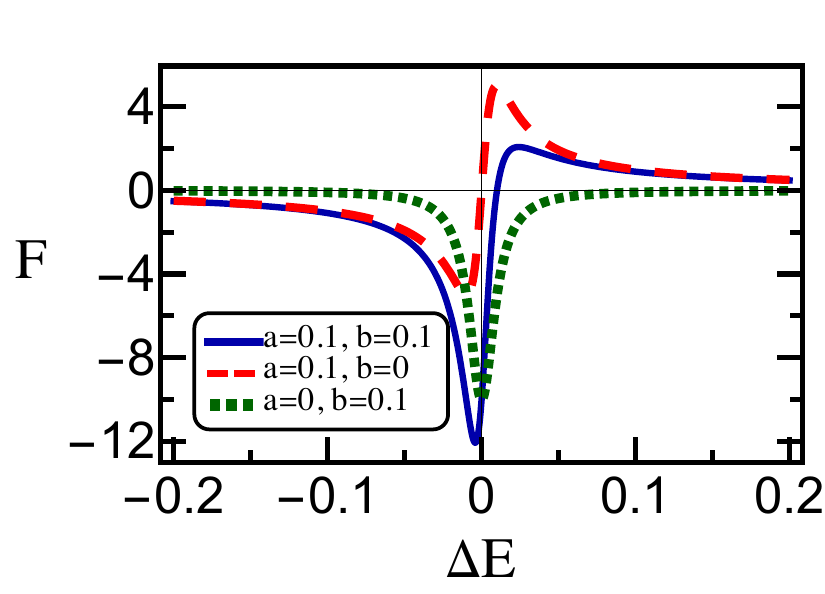}}
		\subfigure []
		{\includegraphics[width=0.4\linewidth,height=0.3\linewidth]{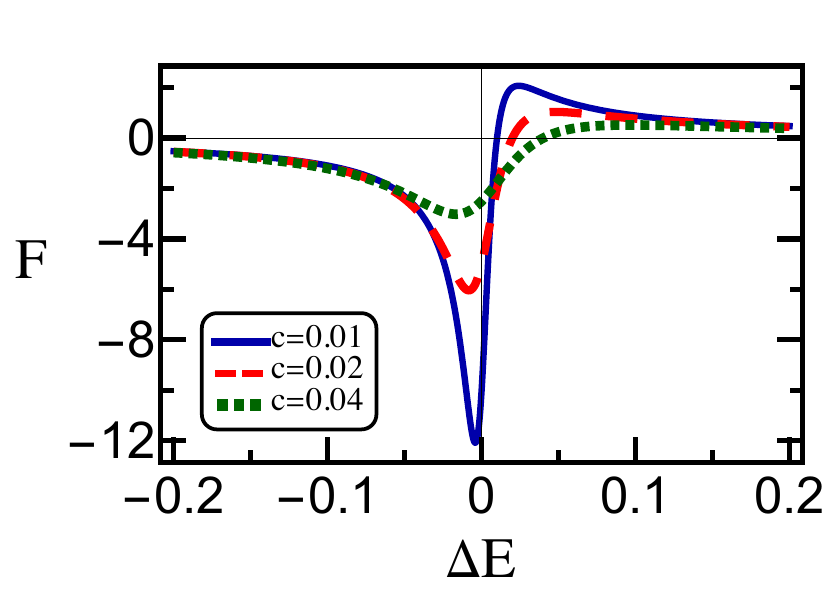}}
		\caption{Variation of the function defined in eq. \eqref{Fano resonance form} with $\Delta E$ for various values of \textbf{(a)} $a,b$ and \textbf{(b)} $E^{(1)}=c$. The default values of parameters are $a=0.1, b=0.1,E^{(1)}=c=0.01$}
		\label{Function_fig}
	\end{figure}
	
	This function has anomalous behavior near the AEDP i.e. when $E_m \approx E_l$.
	Plotting this function for various system parameter configurations in Fig. \ref{Function_fig}, we observe that this function has a possibility of sudden ascent (descent) when $\Delta E$ is in the neighborhood of zero.  This means that we can observe anomalous transport effects like CC near these points. We observe in Fig. \ref{Function_fig} (a) that if both $a, b \ne 0$, the behavior of this function is not symmetric around $\Delta E=0$. However if $a\ne0, b=0$ the behavior becomes symmetric. Additionally, it is also observed  that the local current changes direction as we cross the AEDP. As can be seen by looking at the functional form in equation \eqref{Fano resonance form} when $a=0$, this function becomes a Lorentzian \cite{Quantum_Wheatstone_Bridge} with the first-order corrections to the energy specifying the variance. In Fig. \ref{Function_fig} (b), we see that if the first-order corrections are small, the elevation in the function is much steeper than when these corrections are large. Since the correction is of the order of system bath coupling, we expect it to be small in the approximations we are working on. As a result, we expect that for parameter-induced additional degeneracies in the system, we can have anomalous effects in the current flow if either $a,b\ne0$. Now, since the values of $a,b$ in equation \eqref{Fano resonance form} depends on the eigenvectors needed for diagonalising $\hat{A}^\dagger$ in \eqref{Lyapunov} and since $\hat{A}^\dagger=\frac{\hat{G}+\hat{R}}{2}+i M$, this means that these values depend on the system Hamiltonian as well as the point of connection to the baths. It has been earlier observed \cite{main_reference,Our_circulation} that for either $a,b \ne 0$, the system must possess some form of asymmetry and just the AEDP is not enough to induce exotic transport phenomenon like CC in the system.  We now discuss the Local Lindblad master equation which is a specific case of the GLME discussed earlier,
	\subsection{Local Lindblad Master Equation (LLME)}
		To utilize the idea proposed in the section \ref{circulation_possible}, we will use the LLME \cite{Breuer_PhysRevE.76.031115,Prosen_2008_PhysRevLett.101.105701,ssh_eigen_2} to analyse our system. For this master equation, the rates are given as follows \cite{Lyapunov_PhysRevB.107.035113,Lyapunov_PhysRevE.94.032139},		\begin{align}\label{rate_bath}
			&\gamma_i=\kappa_i J(\Omega)\left(1-\mathcal{N}(\Omega,T_i)\right),
			&&\Gamma_i=\kappa_i J(\Omega) \mathcal{N}(\Omega,T_i)
		\end{align}
		where, $i\in\{1,N_{U+2}\}$ and all other rates are zero. Also, as earlier discussed  $\kappa_{1(N_U+2)}$ is proportional to the square of the interaction strength between system and the left (right) bath, $(\kappa_1 \propto |g^L_K|^2)$.
		Further, we also stick to flat spectrum for the baths so  $J(\Omega)=1$, and the Fermi-Dirac distribution is given as,
		\begin{align}
			\mathcal{N}(\Omega,T_i)=\frac{1}{e^{{\Omega}/{T_i}}+1}
		\end{align}
		where we assume that both the baths are at same chemical potential which can be set to `0' without loss of generality. In the following sections, we study the application of these ideas to two models detailing the conditions for obtaining CC in the systems along with specific system asymmetries required.

	\begin{figure}[t]
		\centering 
		\subfigure []
		{\includegraphics[width=0.4\linewidth,height=0.25\linewidth]{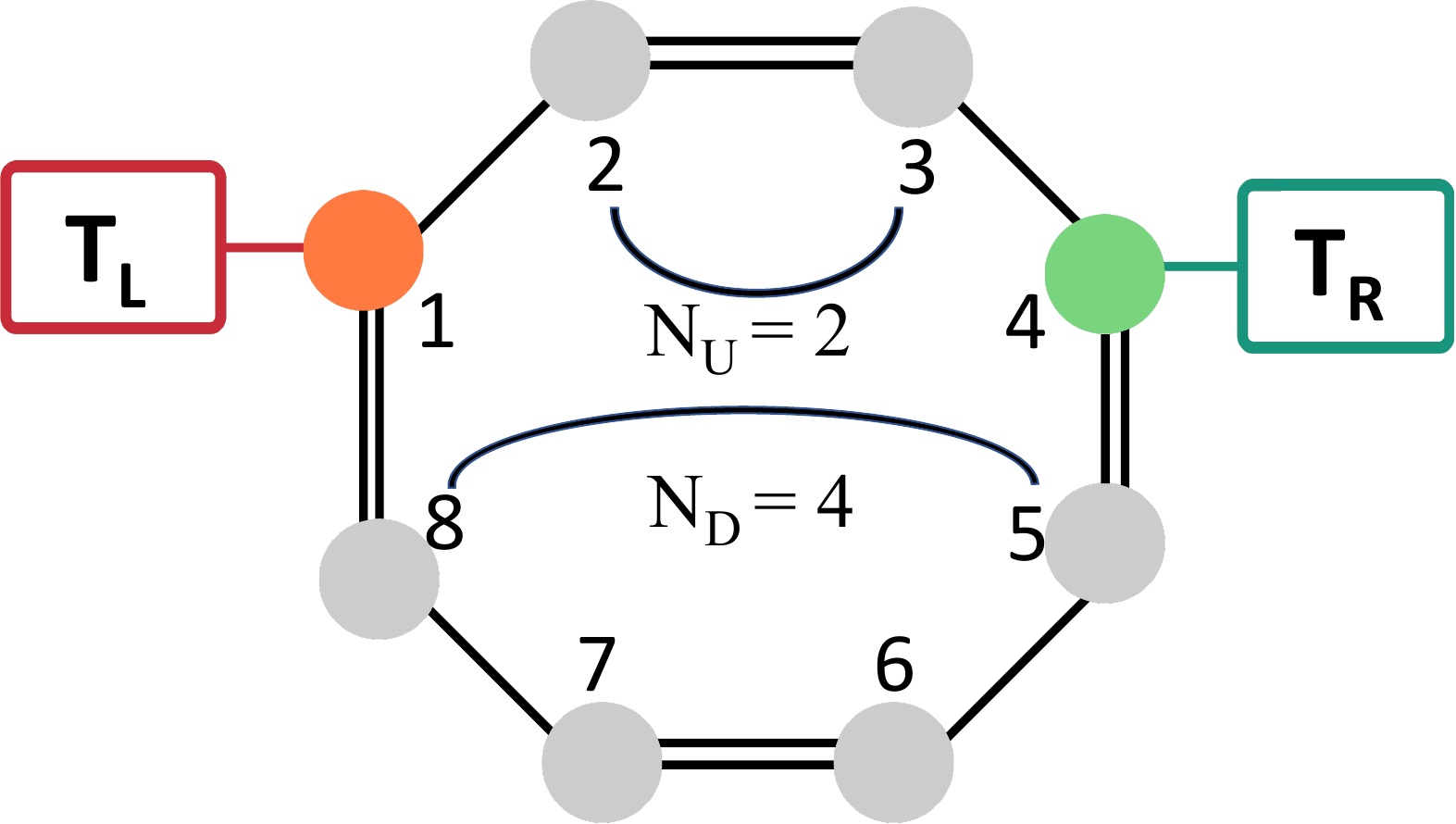}}
		\subfigure []
		{\includegraphics[width=0.4\linewidth,height=0.25\linewidth]{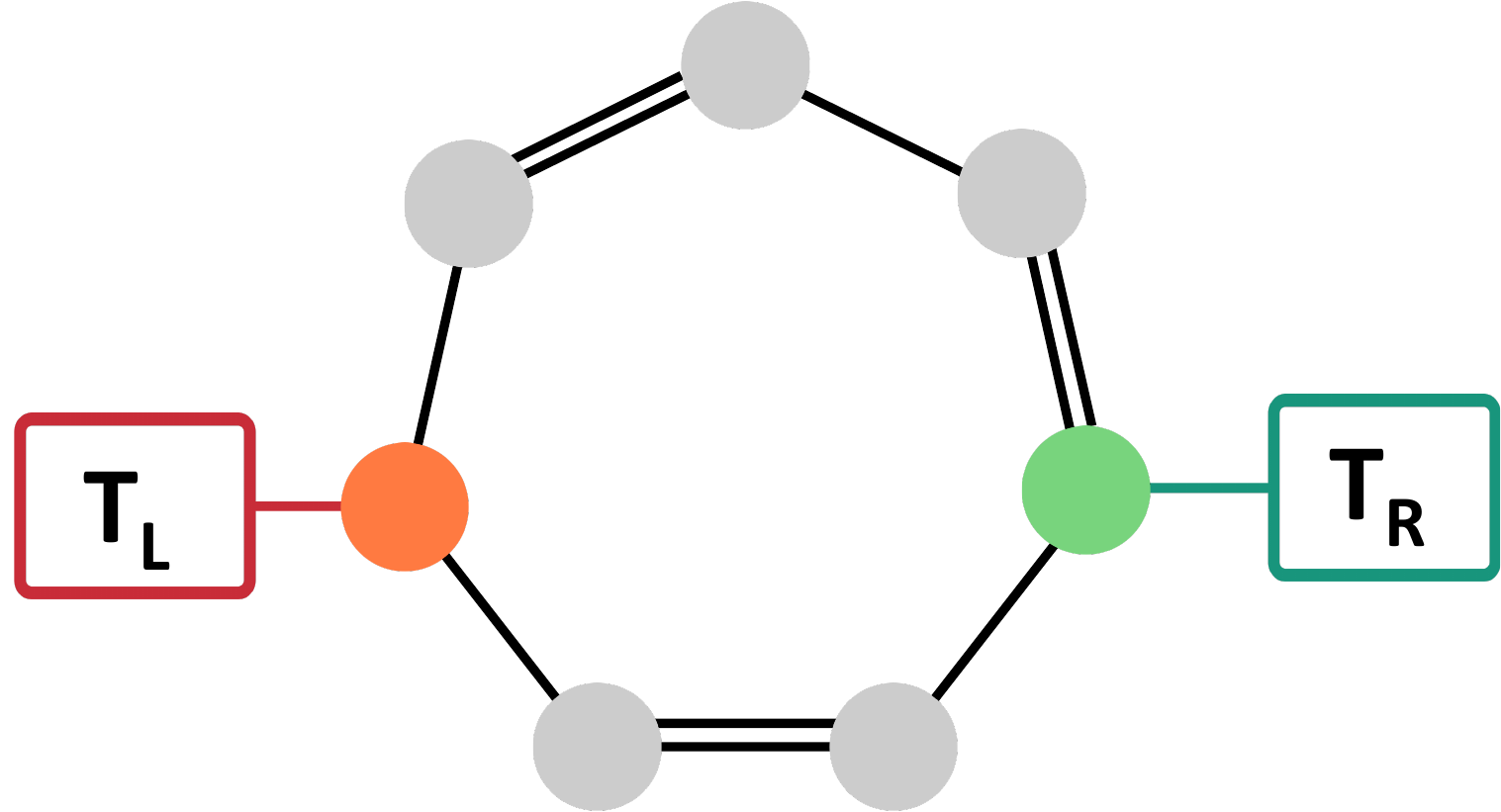}}
		\caption{Schematic  diagram of \textbf{(a)} SSH model with even N and \textbf{(b)} SSH model with odd N with periodic boundary condition.}
		\label{SSH_Model_fig}
	\end{figure}
	\section{Application to the Su–Schrieffer–Heeger (SSH) Model}\label{SSH_Model}
	We begin our application part by considering the SSH model with periodic boundary conditions (see Fig. \ref{SSH_Model_fig}). The Hamiltonian for this system  is given as 	\cite{Ssh_eigen,ssh_original_paper}:
	\begin{align}\label{model_hamiltonian_ssh}
		\hat{H}_S&=t\sum_{n=1}^{N} (1+(-1)^n \delta) (\hat{c}_{n+1}^\dagger \hat{c}_n+\hat{c}_{n}^\dagger \hat{c}_{n+1})+\Omega \sum_{n=1}^N \hat{c}_{n}^\dagger \hat{c}_n
	\end{align}
	where, $\delta$ indicates the difference in hopping strengths of neighboring sites and we follow our earlier convention \cite{Our_circulation} (see Fig. \ref{SSH_Model_fig}(a)) in numbering the Fermionic sites, where $N_U (N_D)$ are the number of Fermionic sites in upper and lower branch respectively which are not in contact with the baths. The numbering is clockwise which means that the Fermionic sites numbered $1$ and $N_U+2$ are in contact with the left and right bath respectively and the total system size $N=N_U+N_D+2$.

	
	\subsection{Particle and Heat Current definition}
	To find the operators corresponding to the steady state heat and particle current in the system, we use the dynamical equation of the density matrix \eqref{local_master_equatio}.	The particle number operator at a site $m$ is given as,
	\begin{align} \label{particle_operator}
		\hat{P}_m=\hat{c}_m^\dagger \hat{c}_m
	\end{align}
	The particle current operator is then defined using the following equation \cite{Prosen_2008_NJP},
	\begin{align}
		\frac{d\langle \hat{P}_m \rangle}{dt}=\langle -i[\hat{P}_m,\hat{H}_S] \rangle=-\langle \hat{J}_m\rangle+\langle \hat{J}_{m-1} \rangle
	\end{align}
	The particle current operator is given as,
	\begin{align}
		\hat{J}_m= i t(1+(-1)^m\delta)  \big{(}\hat{c}_{m+1}^\dagger \hat{c}_m-\hat{c}_{m}^\dagger \hat{c}_{m+1}\big{)}
	\end{align}
	So the average particle current is defined as \cite{ssh_cur_definition,nature_main_ref},
	\begin{align}
		\textnormal{J}_m=	\langle \hat{J}_m \rangle =\langle i  t(1+(-1)^m\delta )\big{(}\hat{c}_{m+1}^\dagger \hat{c}_m-\hat{c}_{m}^\dagger \hat{c}_{m+1}\big{)}\rangle
	\end{align}
	This definition of local current is only valid for the Fermionic sites which are not in direct contact with the baths \cite{Prosen_2008_NJP}.
	Similarly, the local  heat current is found by taking the local bond energy term from the  Hamiltonian in Eq. \eqref{model_hamiltonian_ssh}  \cite{Prosen_2008_NJP}, 
	\begin{align}
		\hat{H}_m&=t(1+(-1)^m\delta) (\hat{c}_{m+1}^\dagger \hat{c}_m+\hat{c}_{m}^\dagger \hat{c}_{m+1})+\Omega  \hat{c}_{m}^\dagger \hat{c}_m
	\end{align}
	This definition of the local energy term is slightly different from the conventional definitions \cite{Prosen_2008_NJP}, which usually involves chemical potential energies from the both sites $m$ and $m+1$. However, this does not lead to any qualitative or quantitative difference in the results we observe. The average current can be found by the following continuity equation:
	\begin{align}
		\frac{d\langle\hat{H}_m \rangle}{dt}=\langle -i[\hat{H}_m,\hat{H}_S] \rangle=-\langle \hat{Q}_m\rangle+\langle \hat{Q}_{m-1} \rangle
	\end{align}
	We do the calculation and find that the average heat current is given as,
	\begin{align}
		\textnormal{I}_m=	\langle\hat{Q}_m\rangle &= \langle it^2(1-\delta^2) \big{(}\hat{c}_{m+2}^\dagger \hat{c}_m-\hat{c}_{m}^\dagger \hat{c}_{m+2}\big{)}+i \Omega t(1+(-1)^m\delta)  \big{(}\hat{c}_{m+1}^\dagger \hat{c}_m-\hat{c}_{m}^\dagger \hat{c}_{m+1}\big{)}\rangle 
	\end{align}
	If `$m$' lies on the upper (lower) branch we call the current $\textnormal{I}_U (\textnormal{I}_D)$.
	Finally, the total current is the sum of the branch currents and according to the convention of this manuscript is given as,
	\begin{align}
		\textnormal{I}_L=I_U-I_D.
	\end{align}	
	The negative sign above comes because of the convention. This is because $I_U$ and $I_L$ are positive when current flows from left to right whereas $I_D$ is negative for this scenario. This means that the current circulates when the signs of $I_U$ and $I_D$ are the same. We see from the definitions of local heat and particle current above that while the particle current only depends on the nearest neighbor correlations, the heat current depends on both the nearest and next to nearest neighbor correlations. This can be understood by realising that while the particle hopping only depends on whether the next site is vacant,  heat transfer happens through the change in bond energy and a hoping Fermion changes the energies of two bonds, so the next to nearest neighbor correlations are present. However, for systems where non-zero correlation is only present between nearest neighbors, heat current and particle current are qualitatively similar and differ by only a multiplying factor. This means that for such models, the onset points of CC are the same for both particle and heat currents, whereas if this is not the case then the onset points  may differ. We will later see in this manuscript examples of both these effects. 
	\\Finally, CC occurs when one of the branch currents is greater than the total current flowing between the system and the bath. We define  $I_C, J_C$ as the difference between the larger branch current and the total current. So it is a measure of the excess current flowing in the branches.
	\begin{align}
		I_C=
		\begin{cases}
			0&|\textnormal{I}_L|-\textnormal{I}_G>0, \\
			\textnormal{I}_{G}-\textnormal{I}_L&|\textnormal{I}_L|-\textnormal{I}_G<0.
		\end{cases}
	\end{align}
	where $\textnormal{I}_G =Max[|\textnormal{I}_U|,|\textnormal{I}_D|]$. Similarly, we define the particle circulating current $J_C$.
	
	\subsection{Even N Results}
\begin{figure}[t]
	\centering 
	\subfigure []
	{\includegraphics[width=0.32\linewidth,height=0.25\linewidth]{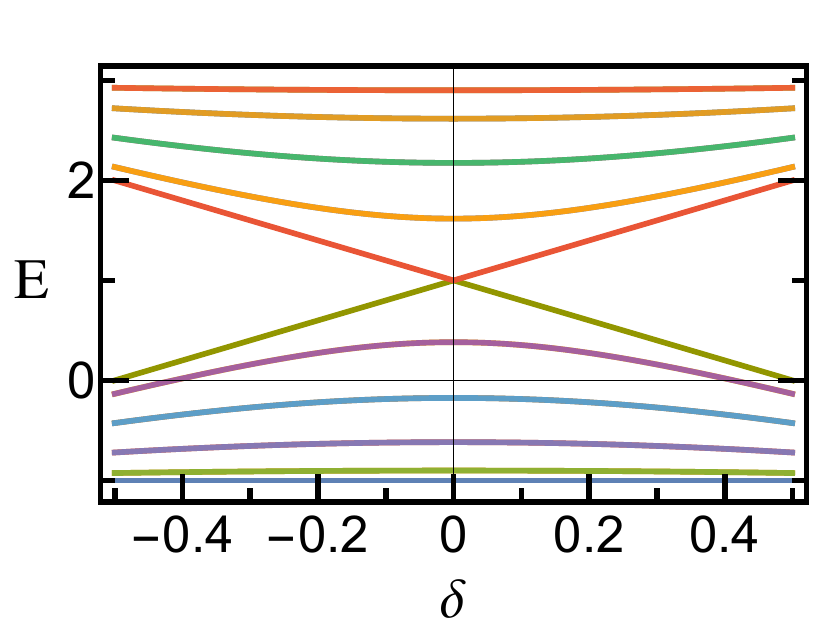}}
	\subfigure []
	{\includegraphics[width=0.32\linewidth,height=0.25\linewidth]{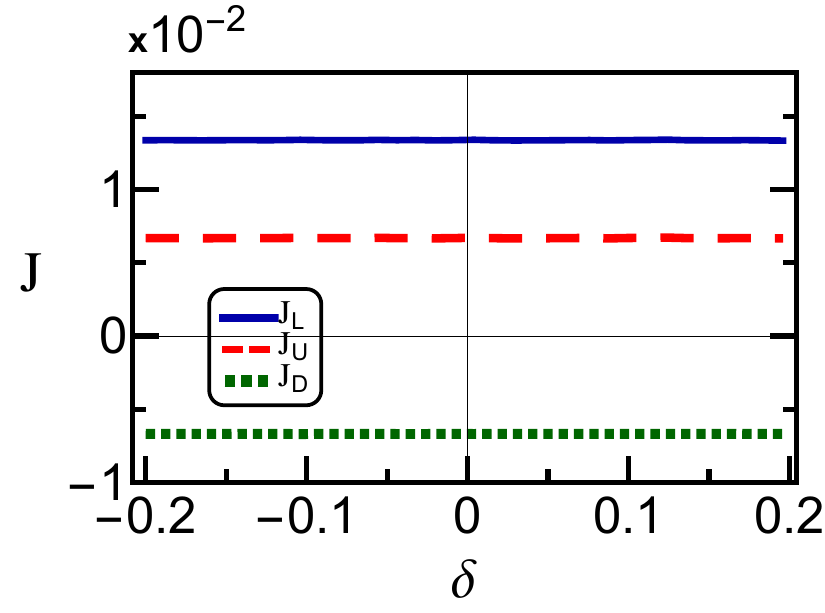}}
	\subfigure []
	{\includegraphics[width=0.32\linewidth,height=0.25\linewidth]{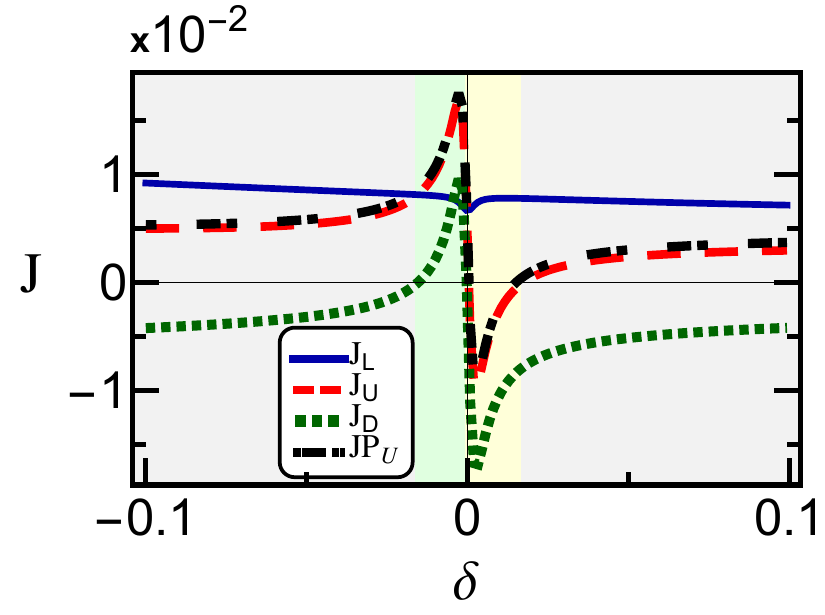}}
	\subfigure []
	{\includegraphics[width=0.32\linewidth,height=0.25\linewidth]{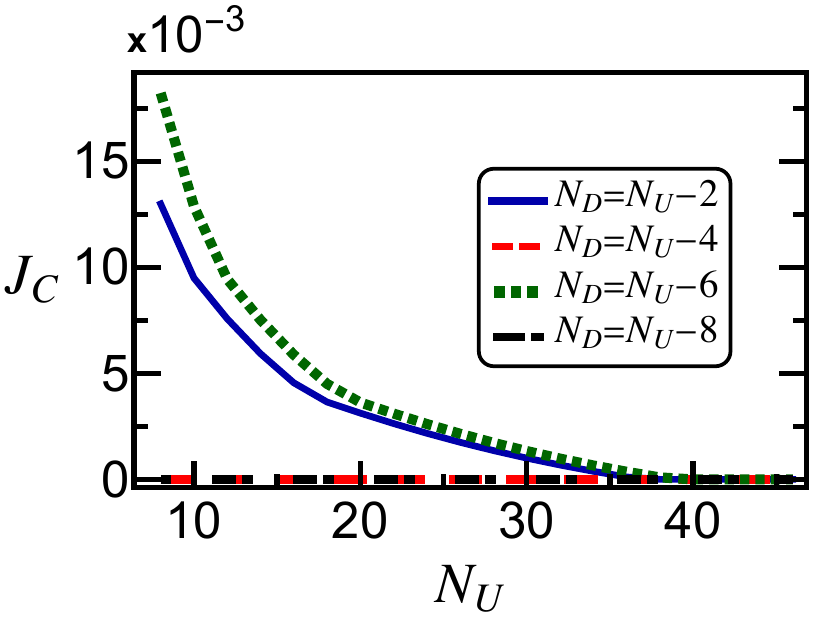}}
	\subfigure []
	{\includegraphics[width=0.32\linewidth,height=0.25\linewidth]{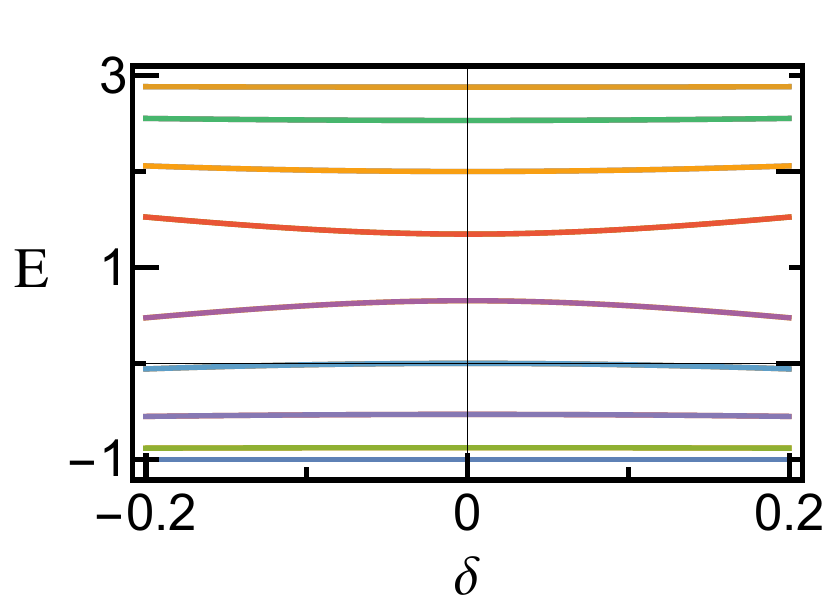}}
	\subfigure []
	{\includegraphics[width=0.32\linewidth,height=0.25\linewidth]{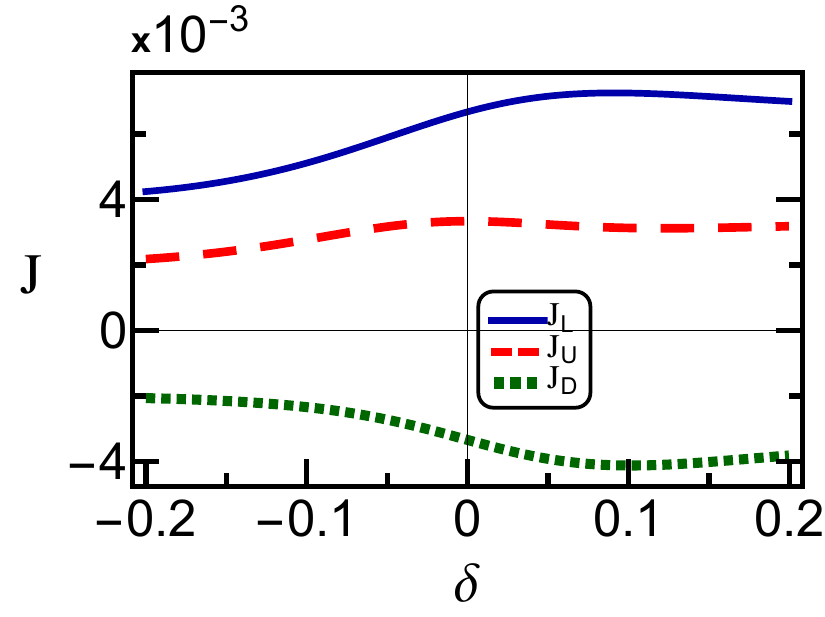}}
	\caption{\textbf{(a)} Variation of energy Spectrum of the SSH model with $\delta$,  Variation of particle currents with $\delta$ for \textbf{(b)} $N_U=N_D=9$, \textbf{(c)} $N_U\ne N_D$. The current `$\textnormal{JP}_\textnormal{U}$' here is the perturbation theory current. \textbf{(d)} Variation of max circulating current with $N_U$, while keeping the number asymmetry, $N_U-N_D$ constant. The maximum value of $J_C$ is taken between the range of $\delta \in \{-0.02,0.02\}$ with one step length of $0.001$. \textbf{(e)} Variation of energy Spectrum of the SSH model with $\delta$ for $N=18$ and variation of particle currents with $\delta$ for \textbf{(f)} $N_U=10, N_D=6$.   If not  specified otherwise, $N_U=10, N_D=8, T_L=1, T_R=0.1$, $\kappa_{1(N_U+2)}=0.1,  \Omega=1$, and $t=1$. In all figures where multiple regions of CC directions are present, we use green, gray, and yellow background colors	to indicate clockwise, parallel, and anticlockwise circulating currents, respectively. For all other cases, we use a white background.}
	\label{current_Even_figures}
\end{figure}
	
	We begin our analysis by considering the chain length $N$ as even. This system can be thought of as two  chains with even or odd number of Fermionic sites in opposite topological phases connected between two baths (see Fig. \ref{SSH_Model_fig}(a)). As $\delta$, the topological order parameter  changes sign, the topological behavior of both these chains reverses. The energy eigenvalues for this system can be explicitly calculated by transforming it to the momentum basis using the Fourier transform,
	\cite{Ssh_eigen,ssh_eigen_2},
	\begin{align}
		\lambda_k=\pm \sqrt{2}t\sqrt{1+\delta^2+(1-\delta^2)\cos{\theta_k}}
	\end{align}
	where, $\theta_k$ can be found by solving the  following equation,
	\begin{align}
		&\theta_k=\frac{4 \pi k}{N},&&k=1,2 \ldots N/2
	\end{align}

	Looking at Fig. \ref{current_Even_figures}(a), we see that the spectrum for the SSH model with system size $N=20$ has an additional degeneracy at $\delta=0$. This means that according to our discussions in  subsection \ref{circulation_possible},  there is a possibility for this system to exhibit CC near this point. On plotting the currents with $\delta$ for a system with the same number of Fermionic sites on the upper and lower branch ($N_U=N_D=9$) in Fig. \ref{current_Even_figures} (b), we see that no CC is observed. This tells us that just the point of additional degeneracy is not enough to observe CC and some additional source of asymmetry may be  required. We employ the asymmetry of having different numbers of sites on upper and lower branch $(N_U=10, N_D=8)$ and  observe that  the current does circulate near the AEDP  for our system (see Fig. \ref{current_Even_figures} (c)). We also see that the perturbative analysis in eq. \eqref{current_contribution} matches very well with the exact results.  The window of $\delta$ where CC occurs is very narrow, because the two energy levels that become degenerate at $\delta=0$ sharply diverge when $\delta$ moves away from this point, resulting in a rapid increase of $\Delta E$ in eq. \eqref{Fano resonance form}. Further, studying the behavior of CC with system size $N$ while keeping the number asymmetry `$N_U-N_D$' constant in Fig. \ref{current_Even_figures} (d), we see that CC only occurs for relatively smaller systems. This may be because there is only one additional degeneracy at $\delta=0$, and as we increase the system size, its contribution to the overall current might become insignificant.  Interestingly, we also see that CC does not occur when the number asymmetry is $4,8$ whereas it occurs for number asymmetry $2,6$. To understand why this is so, we plot the spectrum of a system with $N_U=10, N_D=6$ in Fig. \ref{current_Even_figures} (e) and see that this system does not have the additional degeneracy at $\delta=0$ and as a result, the asymmetry in the number of sites in upper and lower branch does not lead to CC in this system (see Fig. \ref{current_Even_figures}(f)). Considering all the above observations, we see that to obtain CC in our system, we  require both  AEDP and some asymmetry within the system. Having only one of these will not give us CC as can be seen in Figs. \ref{current_Even_figures} (b) and (f).  This system has the same values of  heat and  particle currents. This is because  the imaginary part of $\hat{C}_{i,i+2}=0$ for any site $i$.
Notably, when solving the Lyapunov equation for cases with the same number of  Fermionic sites on the upper and lower branches, we observed that the solver was not able to perform the calculation. This was mathematically linked to having an under-determined set of equations and was overcome by specifying the initial condition of the correlation matrix in eq. \eqref{Correlation_Matrix_dynamic}. It was observed that the current remains the same for different set of initial conditions, and we get equal currents in upper and lower branch for equal number of sites in the two branches (see Fig. \ref{current_Even_figures} (b)).
	\begin{figure}[t]
		\centering 
		\subfigure []
		{\includegraphics[width=0.32\linewidth,height=0.25\linewidth]{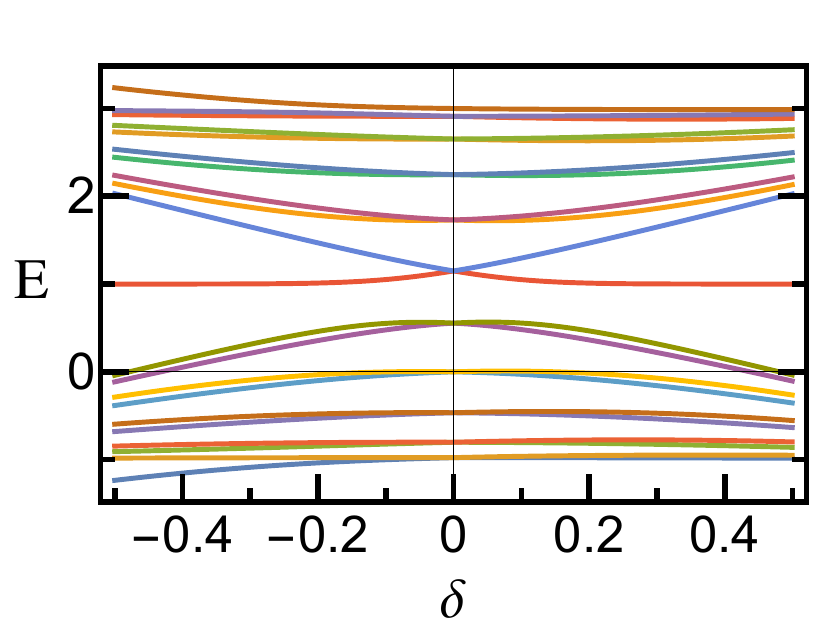}}
		\subfigure []
		{\includegraphics[width=0.32\linewidth,height=0.25\linewidth]{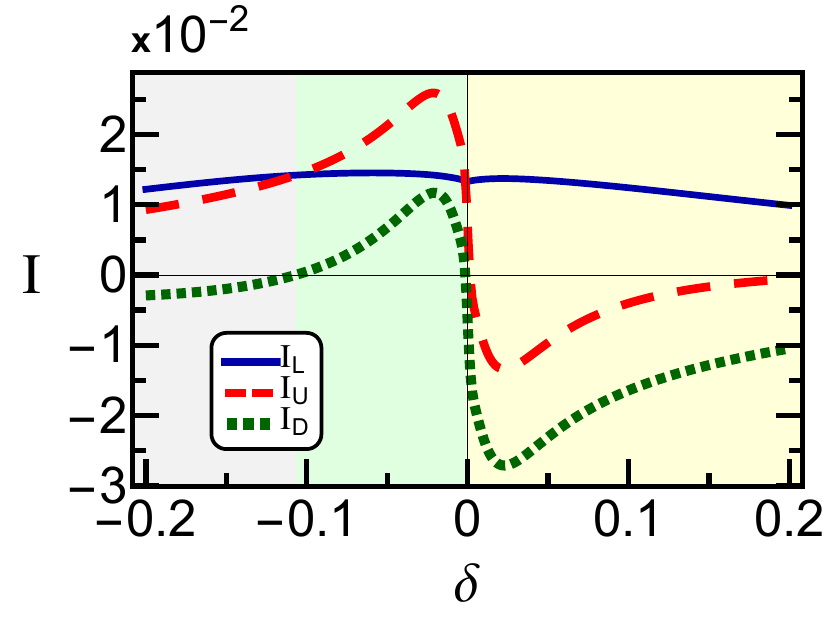}}
		\subfigure []
		{\includegraphics[width=0.32\linewidth,height=0.25\linewidth]{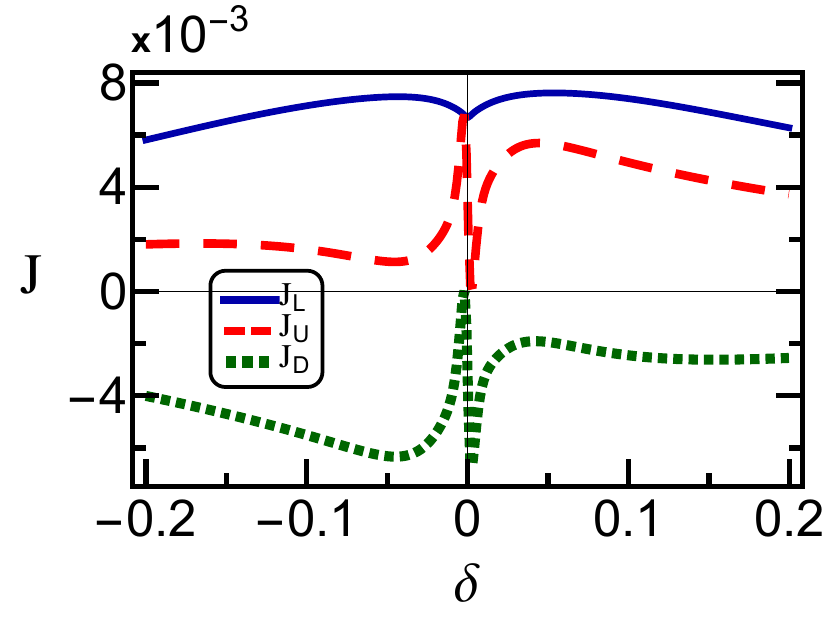}}
		\subfigure []
		{\includegraphics[width=0.32\linewidth,height=0.25\linewidth]{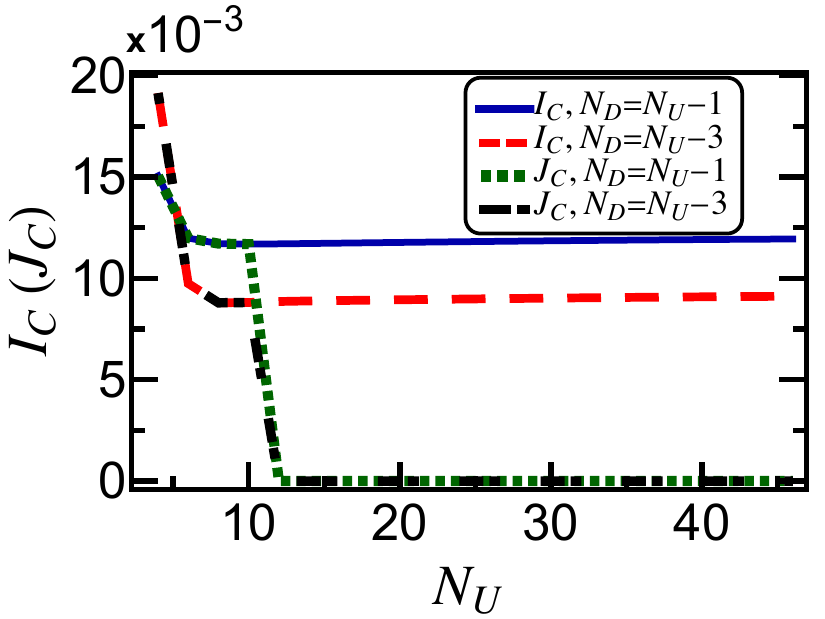}}
		\subfigure []
		{\includegraphics[width=0.32\linewidth,height=0.25\linewidth]{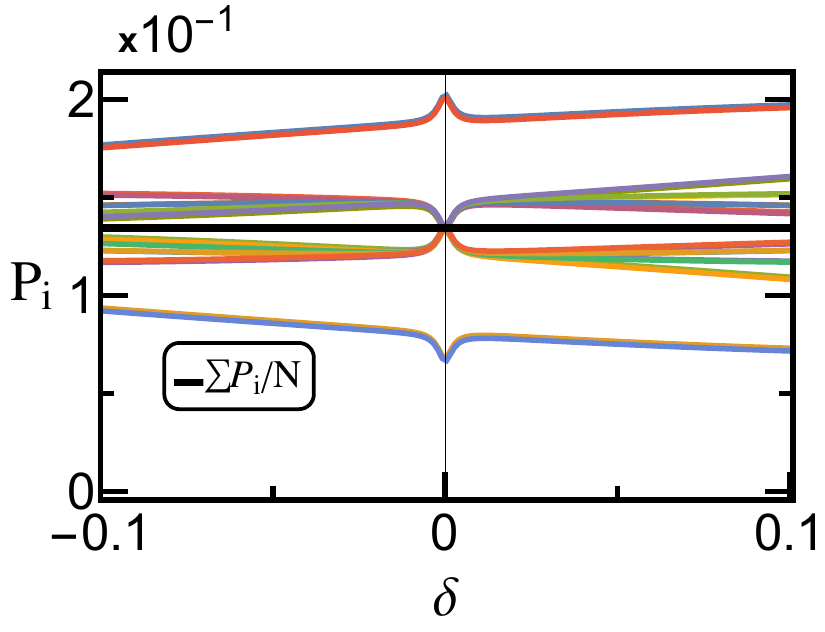}}
		\subfigure []
		{\includegraphics[width=0.32\linewidth,height=0.25\linewidth]{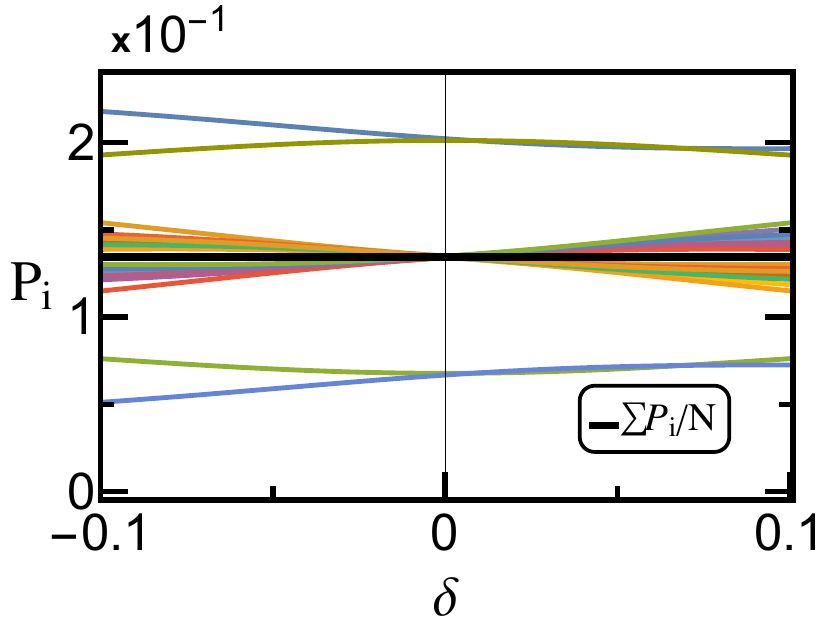}}
		\caption{\textbf{(a)} Variation of energy Spectrum of the SSH model with $\delta$ for odd number of Fermionic sites,  \textbf{(b)} Variation of heat currents with $\delta$,  \textbf{(c)} Variation of particle currents with $\delta$  and \textbf{(d)} Variation of max circulating current with $N_U$, while keeping the number asymmetry $N_U-N_D$ constant. Variation of average occupancy at all sites with $\delta$ for a system with \textbf{(e)} $N_U=10, N_D=8$ and \textbf{(f)} $N_U=10, N_D=6$
			If not  specified otherwise, $N_U=10, N_D=9$.}
		\label{current_Odd_figures}
	\end{figure}

	\subsection{Odd N}\label{subsection:Odd}
	We now consider systems having size $N$ odd (see Fig. \ref{SSH_Model_fig} (b)). This can be thought of as two SSH chains, one with an even number of Fermionic sites and the other with an odd number of Fermionic sites connected between two baths.  Plotting the energy spectrum for the model with $N=21$ in  Fig. \ref{current_Odd_figures} (a), we observe that the spectrum changes significantly if we have an odd number of Fermionic sites. Though the AEDPs still remains at $\delta=0$, there are many more additional degeneracies at this point instead of just one as in the earlier case. Plotting the corresponding heat and particle current for the setup $N_U=10, N_D=9$ in Fig. \ref{current_Odd_figures} (b), (c) respectively, we see that the behavior of currents is quite different for this model in comparison to the previous case. The value of heat and particle current is not the same, which can be attributed to non-vanishing of imaginary part of  next to nearest neighbor correlations $\hat{C}_{i,i+2}\ne0$ for this model. Also, looking at the heat CC in Fig. \ref{current_Odd_figures} (b), we see that the window of the region where heat CC occurs is much broader than in the previous case. This may be because, in the energy spectrum, many energy levels that become degenerate at $\delta=0$,  do not sharply diverge as we move away from this point, so $\Delta E$ increases slowly as compared to the previous model.  However, this window is still very narrow for the particle CC in Fig. \ref{current_Odd_figures} (c). This tells us that   the support for particle CC is coming only from those degenerate energy levels which diverge sharply as we move away from $\delta=0$. Looking again at the energy spectrum, we see that only two degenerate energy levels diverge sharply, so the particle CC may exist only due to  the interaction between these two energy levels. Finally, plotting the maximum heat and particle CC with $N_U$ while keeping the number asymmetry $N_U-N_D$ constant in Fig. \ref{current_Odd_figures} (d), we observe that the heat CC sustains for larger system sizes. This may be because, unlike the previous case, here the AEDPs do increase with system size and hence the anomalous contributions to the total current remain significant. However, the particle CC does not sustain in the larger systems. This is in line with our earlier assumption that the support for particle CC comes from the interaction of only two degenerate levels and this contribution will become insignificant as the system size becomes large.
	Considering all this, we have an interesting scenario where heat current circulates while particle current runs parallelly in the branches. This is because heat CC has the support of more degenerate energy levels than particle CC. In earlier studies \cite{main_reference,Our_circulation,Quantum_Wheatstone_Bridge}, it  has also been observed that in the systems with presence of 
	anomalous current behavior near certain points, other physical 
	observables also can have have non-trivial behavior near that point. To check whether this is true for our system, we plot the spectrum of average site occupancy $P_i=\hat{C}_{ii}$ with $\delta$ for all sites in Fig. \ref{current_Odd_figures} (e) and (f). We observe that for system with presence of AEDP and CC ($N_U=10, N_D=8$, Fig. \ref{current_Odd_figures} (e)), the behavior of average site occupancy is non-trivial near AEDP, whereas for system with no AEDP and CC ($N_U=10, N_D=6$, Fig. \ref{current_Odd_figures} (f)), the variation of average site occupancy is rather smooth and typical. In both the cases, we also see that the average total number of particles in the system $\sum_i \hat{C}_{ii}/N$ remains almost unperturbed by a change in the value of $\delta$.
	
	\section{Comparison between exact and Local Lindblad Master Equation (LLME) results}\label{NEGF_results}
	 \begin{figure}[b]
		\centering 
		\subfigure []
		{\includegraphics[width=0.32\linewidth,height=0.25\linewidth]{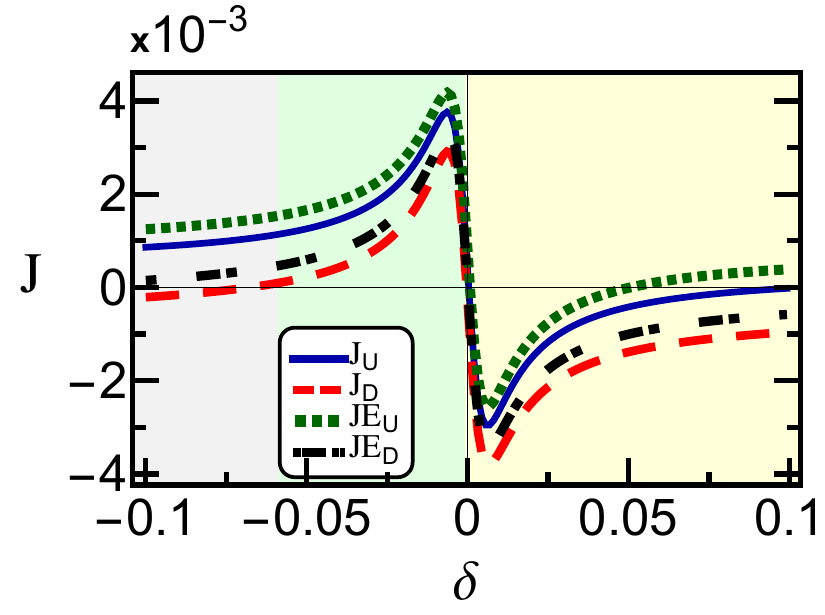}}
		\subfigure []
		{\includegraphics[width=0.32\linewidth,height=0.25\linewidth]{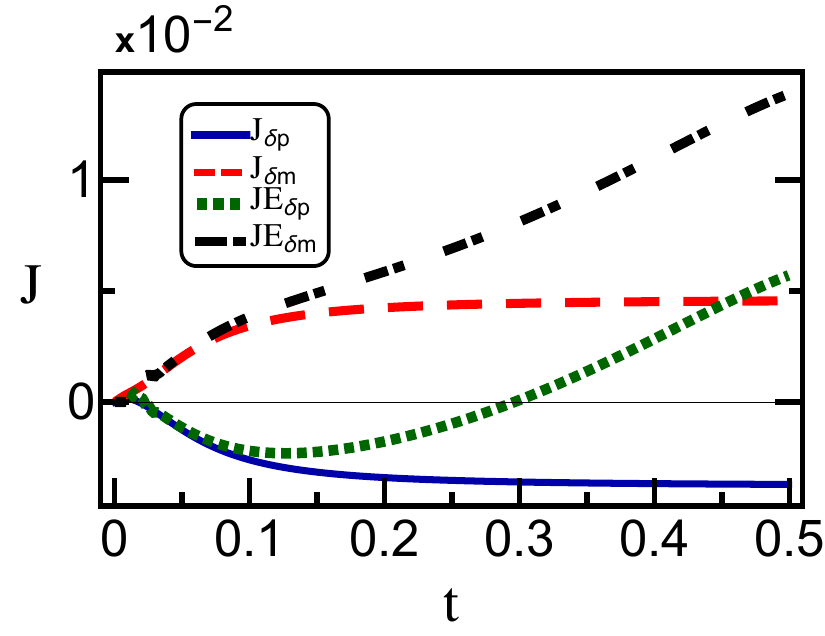}}
		\subfigure []
		{\includegraphics[width=0.32\linewidth,height=0.25\linewidth]{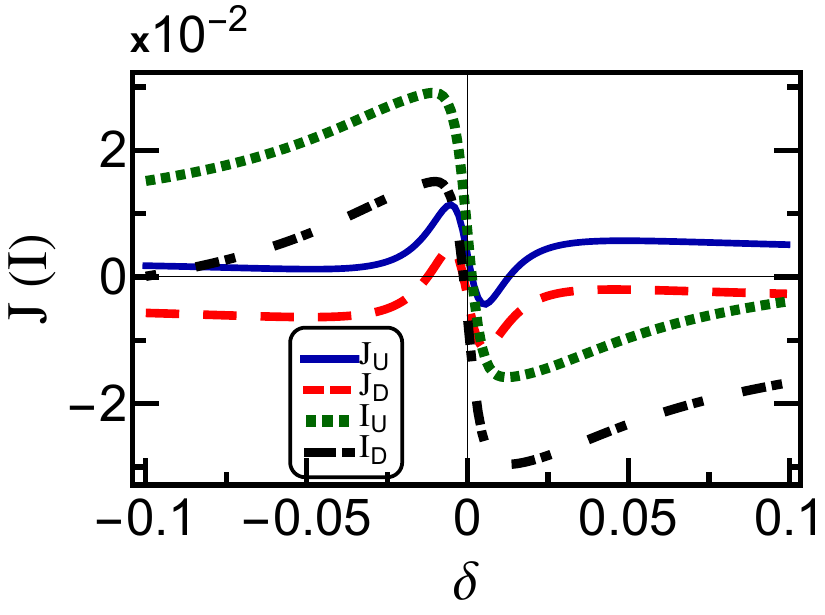}}
		\subfigure []
		{\includegraphics[width=0.32\linewidth,height=0.25\linewidth]{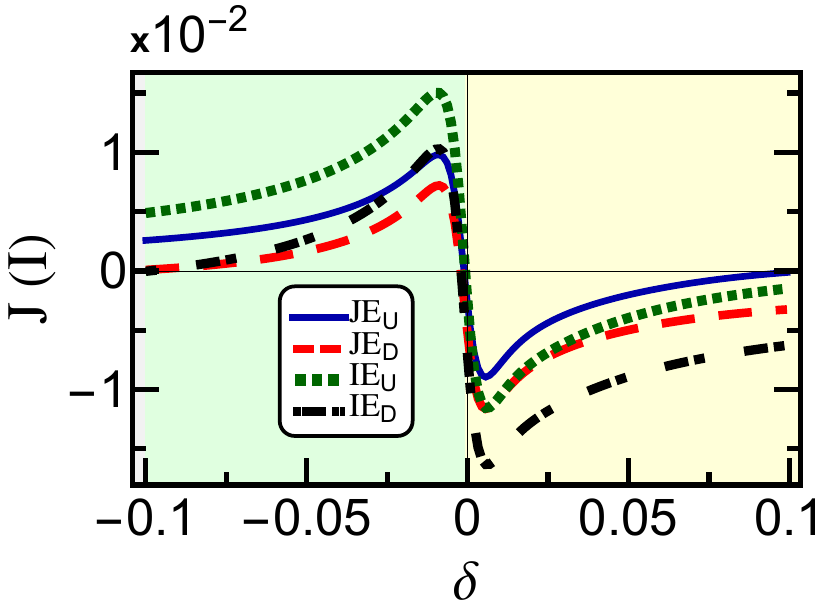}}
		\subfigure []
		{\includegraphics[width=0.32\linewidth,height=0.25\linewidth]{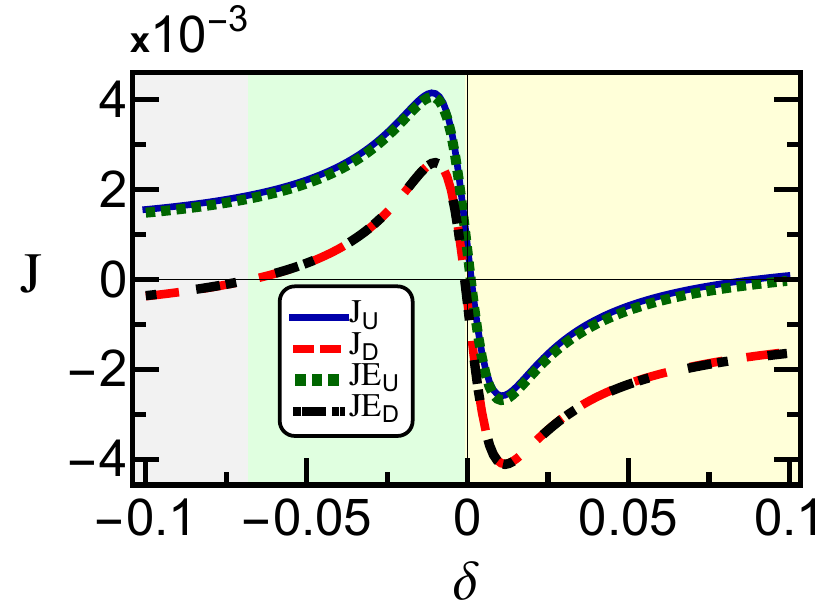}}
		\caption{\textbf{(a)} Variation of branch currents with $\delta$ using the two formalism, `JE' signifies the currents obtained using NEGF formalism with J$\text{E}_\text{{U(D)}}$ signifying the corresponding upper (lower) branch current, \textbf{(b)} Variation of currents with the intersite hopping strength $t$ with the SSH asymmetry parameter, $\delta=\delta_p=0.01, \delta=\delta_m=-0.01$.  Variation of particle and heat current for the for the setup $N_U=4, N_D=3$ with $t=1, \kappa_1=\kappa_2=0.1$, using \textbf{(c)} LLME formalism  and \textbf{(d)}  NEGF formalism.  \textbf{(e)} Variation of branch currents using LLME and NEGF for the bath spectrum function 	$J(\omega)=2\sqrt{1-\frac{\omega^2}{4}} $. 	If not  specified otherwise, $N_U=4, N_D=2, T_L=1, T_R=0.1$, $\kappa_{1(N_U+2)}=0.01,  \Omega=1., t=0.1, \omega_c=4$} 
		\label{Comp_figures}
	\end{figure}
	To check the reliability of the approximate results obtained via the LLME formalism, we compare our approximate results with the exact results obtained via the Non Equilibrium Green Function (NEGF) formalism described in detail in the study \cite{Dhar_NEGF_PhysRevB.73.085119}. According to this formalism, the exact value for the currents for non interacting Fermionic Hamiltonians can be obtained by evaluating the Green function \cite{Dhar_NEGF_PhysRevB.73.085119},
	\begin{align}
		G^{+}(\omega)=\frac{1}{(\omega-M)I-\Sigma_L^+(\omega)-\Sigma_R^+(\omega)}
	\end{align}
	where, $\Sigma_{L(R)}^+(\omega)$ is the self energy operator corresponding to the left (right) bath. Since only the site numbered $1$ $(N_u+2)$ of the system is connected to the left (right) bath, all elements of the left (right) self energy operator are zero except $\Sigma_{L}^+(\omega)_{1,1}$  $(\Sigma_{R}^+(\omega)_{Nu+2,Nu+2})$. Now, according to the NEGF formalism \cite{Bijay_PhysRevLett.130.187101}.
	\begin{align}
		\Sigma_{L}^+(\omega)_{1,1}=\kappa_1 \int \frac{d\omega' J(\omega')}{2 \pi (\omega-\omega')}-i\kappa_1 \frac{J(\omega)}{2}
	\end{align}
	And similar expression can be found for the right self energy operator.  We have earlier assumed that the baths have flat spectrum. Suppose, there is a upper limit to the allowed frequencies $\omega_c$, such that $\omega_c>\Omega$. This means that the spectrum function is given as,
	\begin{align}
		J(\omega)=
		\begin{cases} 1,&0<\omega\le \omega_c
			\\0,&\omega>\omega_c
		\end{cases}	
	\end{align}
	Using this the left self energy element becomes, for $\omega<\omega_c$,
	\begin{align}
		\Sigma_{L}^+(\omega)_{1,1}=\frac{\kappa_1 }{2 \pi}\log\big{(} \frac{\omega}{(\omega_c-\omega)}\big{)}- \frac{i}{2}
	\end{align}
	Finally, the correlation matrix element in the steady state evaluated according to the NEGF formalism are given as \cite{Dhar_NEGF_PhysRevB.73.085119},
	\begin{align}
		\hat{C}_{i,j}=\int d\omega [(G^+(\omega) \xi^L G^-(\omega))_{ji}N(\omega,T_L)+(G^+(\omega) \xi^R G^-(\omega))_{ji}N(\omega,T_R)]
	\end{align}
	where $\xi_{L (R)}$ are the density matrices defined as $\xi^{L}_{11}=\kappa_1\frac{J(\omega)}{2\pi}$, and $\xi^{R}_{Nu+2,Nu+2}=\kappa_2\frac{J(\omega)}{2\pi}$. All other elements of both the matrices are zero. 

	\par On comparing the values of the currents using LLME and NEGF method, we observe in Fig. \ref{Comp_figures} 
  (a) that for the system where the intersite hopping strength $t$ is much smaller that the onsite chemical potential $\Omega$, the two formalism give almost the same results. The local branch currents changes directions near the point $\delta=0$. This is consistent with the regime of the validity of the LLME \cite{Breuer_PhysRevE.76.031115}. Further,  in Fig. \ref{Comp_figures} (b), we see that on increasing the value of $t$, the values of current obtained via the the two formalisms start to differ. Examining the results for large values of $t=\Omega=1$, we see in fig. \ref{Comp_figures} (c) that the behavior of the particle and heat current is drastically different using the LLME formalism and the heat current has much wider window of circulation than the particle current. This however is not true using the NEGF method. In fig. \ref{Comp_figures} (d) it is seen that the two types of current differ only quantitatively and have the same qualitative behavior. However, in both the cases, the currents still changes direction as we cross the $\delta=0$ point. To examine how the currents depend on the bath spectrum, we analyze the variation of the currents with $\delta$, using the following bath spectrum, 
		\begin{align}
			J(\omega)=2\sqrt{1-\frac{\omega^2}{4}},   -2<\omega<2
 		\end{align} This type of spectrum arises when the bath is modeled by an infinitely extended tight-binding model \cite{Dhar_NEGF_PhysRevB.73.085119}. We observe in Fig. \ref{Comp_figures} (e) that while this change affects the current values quantitatively,  the relationship between CC and AEDP remains intact. Additionally, the agreement between the LLME and NEGF currents is much better for this bath. It is also important to note that the LLME used in the present manuscript can be derived using alternate means that have some other microscopic origins, like the collision models \cite{Collision_doi:10.1142/S1230161222500159}, and hence the NEGF formalism given here may not be describing the same microscopic model.
	
	\subsection{Plausible Physical mechanism behind circulation}
\begin{figure}[h]
		\begin{center}
			\includegraphics[width=0.65\textwidth,angle=0]{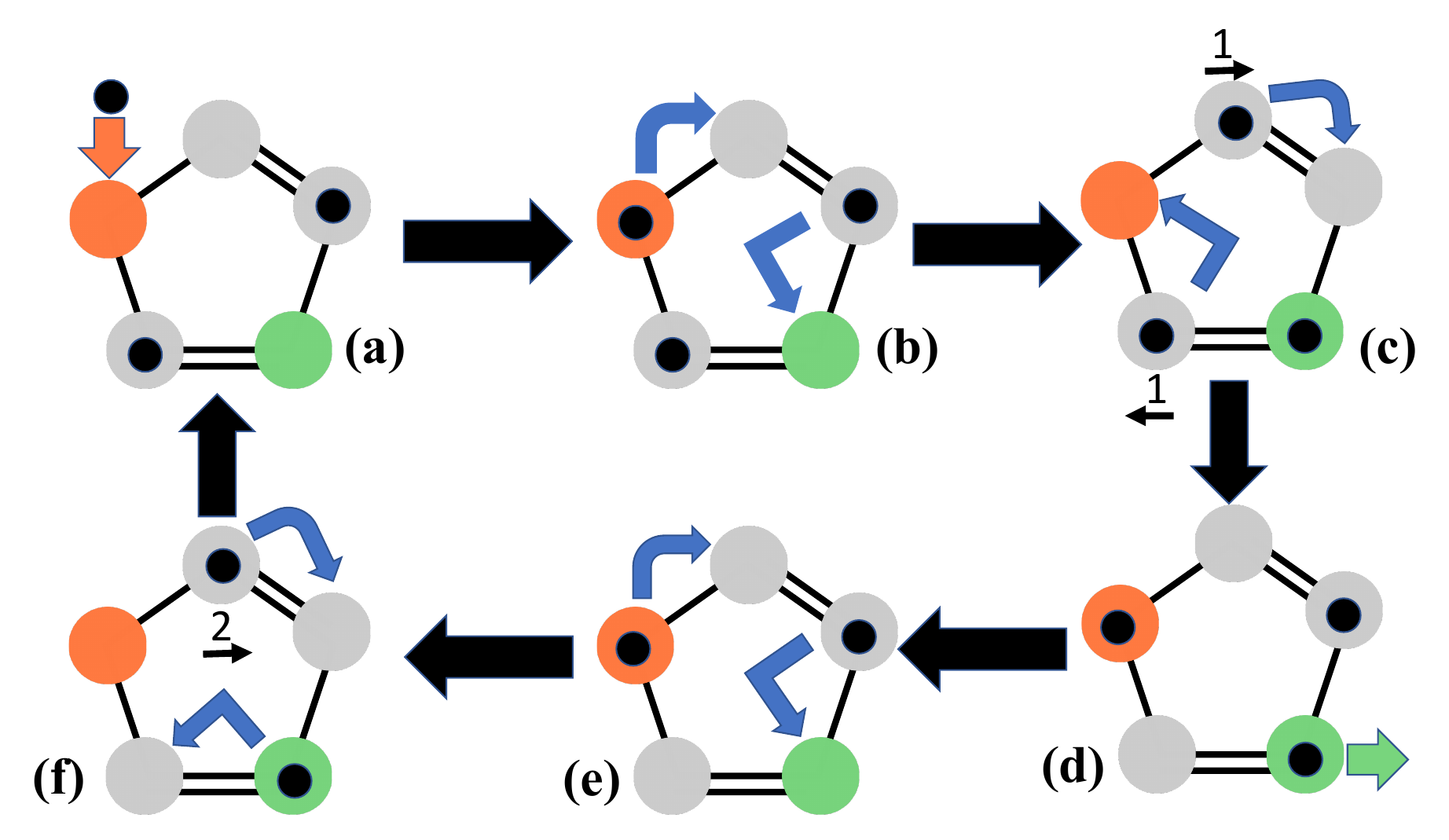}
			\caption{Possible current cycle in a 5-site system resulting in the transfer of 1 particle from left bath to right bath and circulation of a particle in the clockwise direction within the branches.}
			\label{Physica_Mechanism}
		\end{center}
	\end{figure}

	To understand how particle current may circulate in our system, we look at a 6-step cyclic process in a  5-site Fermionic setup $(N_U=2, N_D=1)$  as shown in Fig. \ref{Physica_Mechanism}.  Since, the time evolution of a state is governed by the local master equation \eqref{local_master_equatio}, it means that the baths can only alter the state of sites they are in direct contact with and this process is irreversible \cite{Quantum_Wheatstone_Bridge}.  The rates of these transitions are proportional to the coefficients specified in eq. \eqref{rate_bath}. The change in states of Fermionic sites not in direct contact with the bath is due to the Hamiltonian given in eq. \eqref{model_hamiltonian_ssh} and these changes are reversible in nature. The rates of such transitions depend upon the square of the matrix element of the Hamiltonian and hence depend on the hopping rates in eq. \eqref{model_hamiltonian_ssh}.
	\par Now that we know the evolution rules, let's start from the  initial configuration $| - -\bigcdot-\bigcdot \rangle$, where `$-$' at a position means that the Fermionic site at that position is empty and `$\bigcdot$' means that the site is occupied. As can be seen in Fig. \ref{Physica_Mechanism}, the sites numbered $1$ and $4$ are in contact with the left and right bath respectively. The first step in the process is the transfer of a particle from the left bath to the system ( see Fig. \ref{Physica_Mechanism} (a) ) taking the state from $| - -\bigcdot-\bigcdot \rangle \to | \bigcdot -\bigcdot-\bigcdot \rangle$. This is followed by two reversible steps in Fig. \ref{Physica_Mechanism} (b), (c) where one particle travels left to right from site numbered 2 and one particle travels right to left from site $5$ and we end up with the state $| \bigcdot -\bigcdot \bigcdot -\rangle$. In the next step shown in Fig. \ref{Physica_Mechanism} (d),  one particle exits from site $4$ to the right bath resulting in the transition $| \bigcdot -\bigcdot \bigcdot -\rangle\to | \bigcdot -\bigcdot - -\rangle$.  The final two steps are again carried out by the Hamiltonian and involve the passing of the second particle from left to right at site $2$ and returning to the initial configuration  $| - -\bigcdot-\bigcdot \rangle$ we started with (see Fig. \ref{Physica_Mechanism}(e), (f)). Looking at the cumulative effect of all these processes, we have one particle entering from the left bath and one exiting to the right bath along with the transfer of two particles from left to right in the upper branch, and one particle from right to left  in the lower branch. This means that there is circulation of one particle in the clockwise direction within the branches, and transfer of one particle between the baths through the system. Reversing the sign of transitions happening due to the Hamiltonian, we can have a process resulting in anticlockwise CC in the system. However, due to the irreversibility of processes associated with bath connected sites, the rate of this process will be different than the rate of the clockwise CC process resulting in a net CC.
	\\Similar, to the above process, there  are other processes resulting in CC in the system. They compete with processes having local parallel currents in the branches and the net current behavior is the overall summation of all such processes. Usually, the probability rate for getting parallel transport is much larger than getting CC and hence we observe parallel currents in typical setups. Here,  the rates of processes with CC may be becoming larger than the rates of processes with parallel transport  \cite{Quantum_Wheatstone_Bridge} near the AEDPs, resulting in the observations of overall CC in the system.

	\section{Model with unequal upper and lower branch hopping strengths}\label{Asymm_Model}
	\begin{figure}[h]
		\begin{center}
			\includegraphics[width=0.45\textwidth,angle=0]{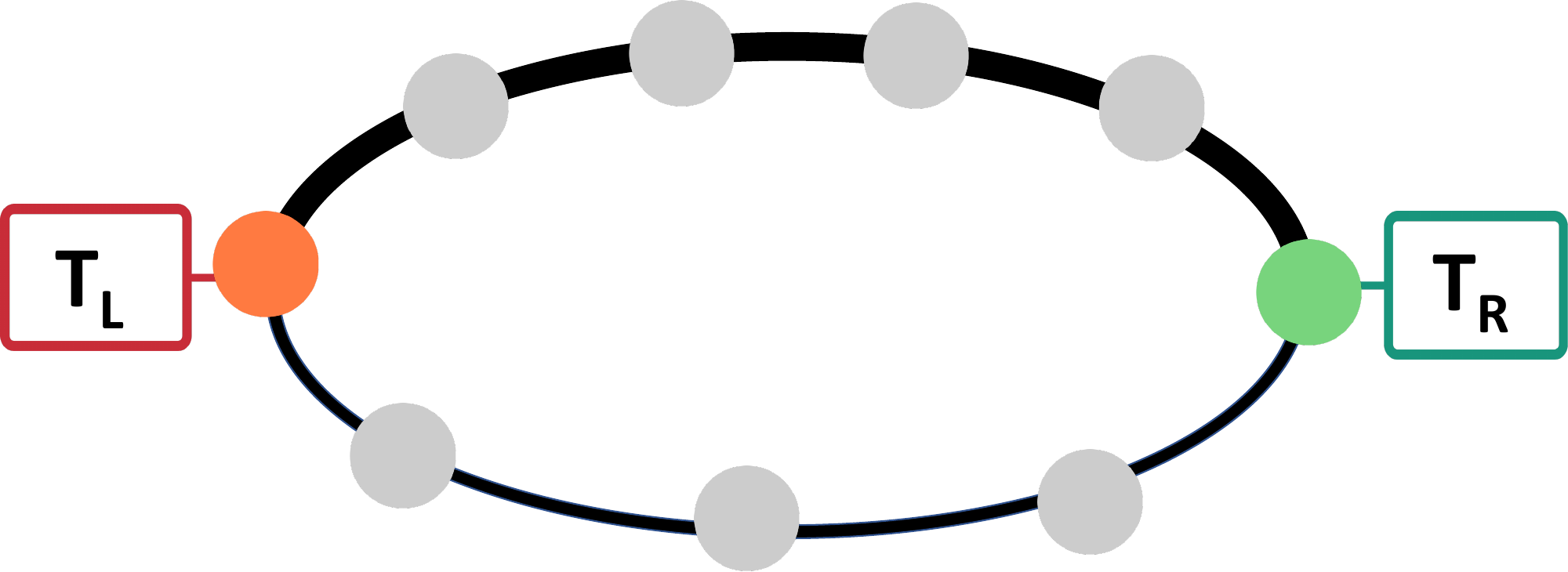}
			\caption{Model with unequal hopping strength in upper and lower branch}
			\label{Asymm_model_fig}
		\end{center}
	\end{figure}
	
	We extend our analysis to another model having unequal hopping  strengths in the upper and lower branch (see Fig. \ref{Asymm_model_fig}). A classical version of a similar model for spins has been studied for CC in the study  \cite{Our_circulation}. The Hamiltonian for this model is given as,
	\begin{align}\label{model_hamiltonian_asym}
		\hat{H}_S&=t\sum_{n=1}^{N} v_n (\hat{c}_{n+1}^\dagger \hat{c_n}+\hat{c}_{n}^\dagger \hat{c}_{n+1})+\Omega \sum_{n=1}^N \hat{c}_{n}^\dagger \hat{c_n}
	\end{align}
		\begin{figure}[t]
		\centering 
		\subfigure []
		{\includegraphics[width=0.4\linewidth,height=0.3\linewidth]{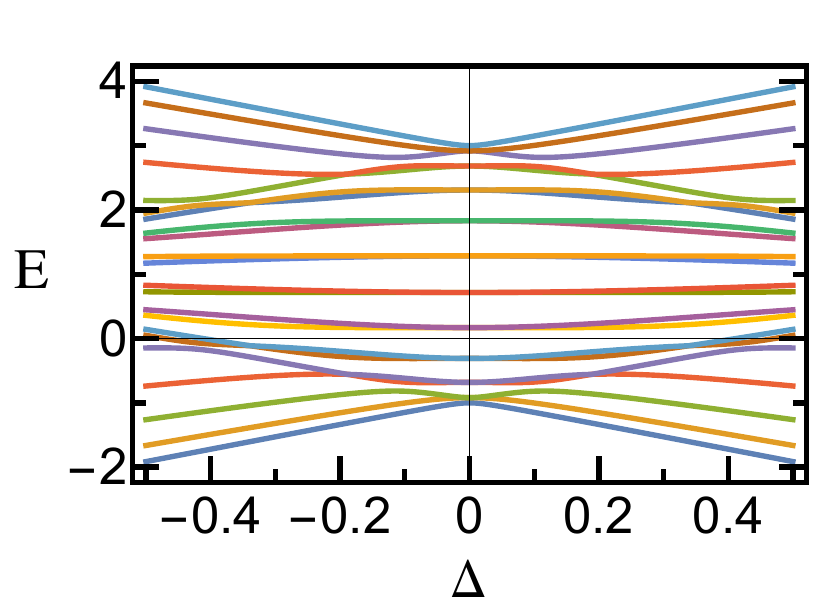}}
		\subfigure []
		{\includegraphics[width=0.4\linewidth,height=0.3\linewidth]{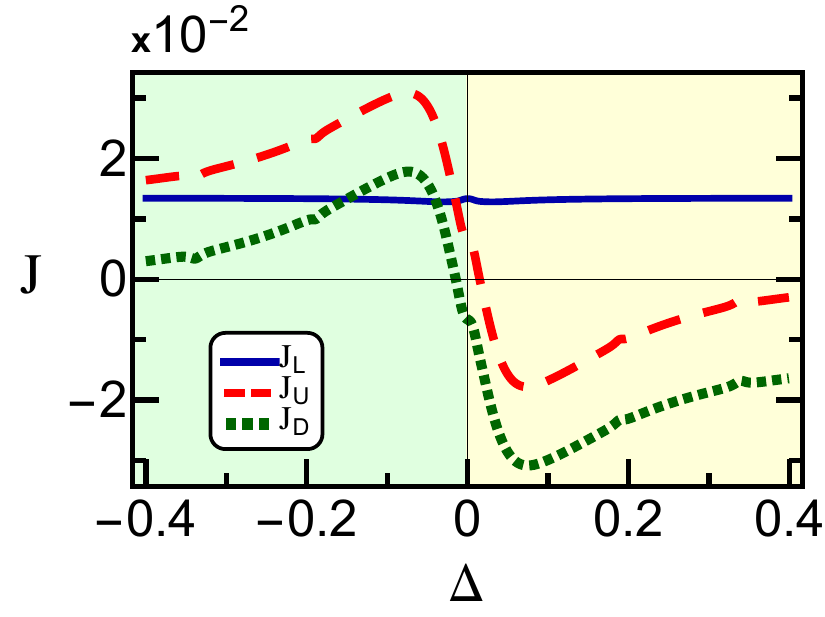}}
		\subfigure []
		{\includegraphics[width=0.4\linewidth,height=0.3\linewidth]{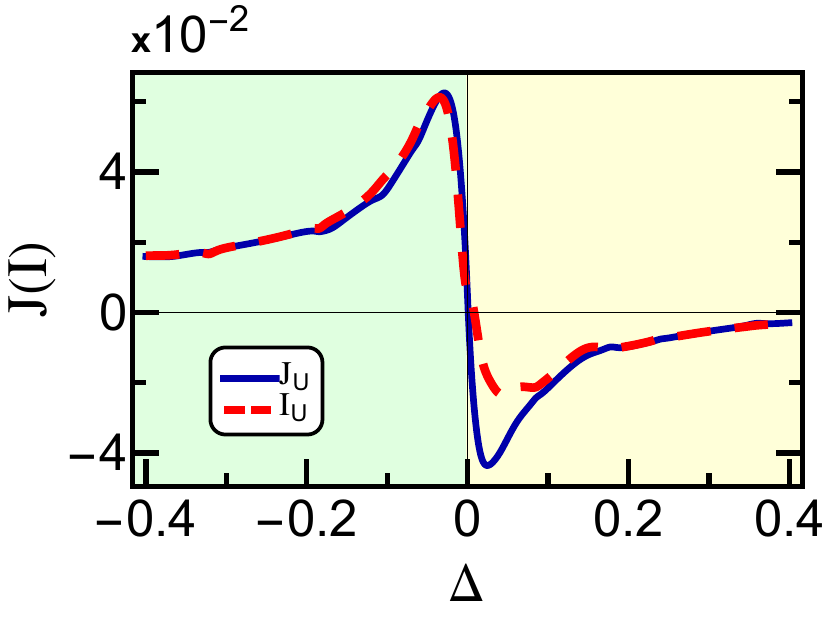}}
		\subfigure []
		{\includegraphics[width=0.4\linewidth,height=0.3\linewidth]{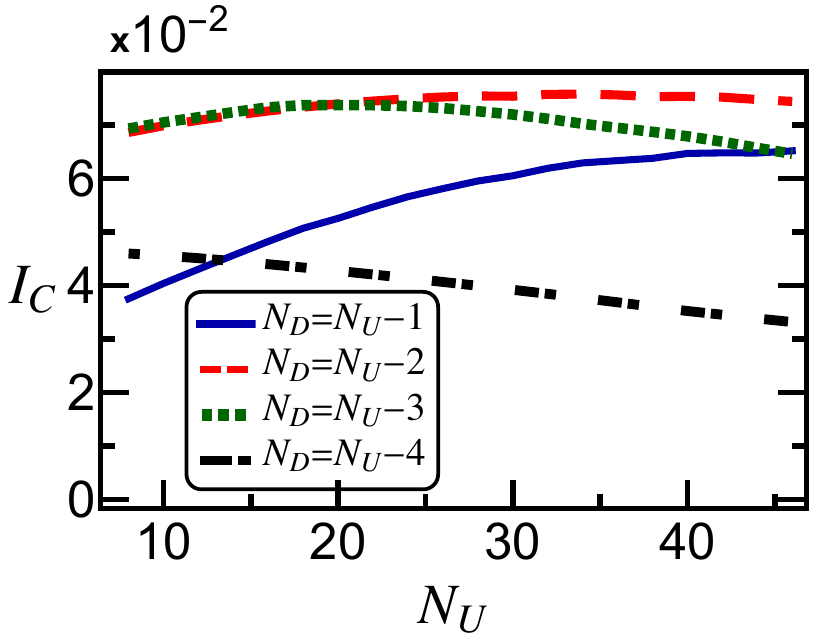}}
		\caption{ \textbf{(a)} Variation of energy Spectrum of the asymmetric model with $\Delta$,  \textbf{(b)} Variation of heat currents with $\Delta$,  \textbf{(c)} Variation of up branch heat and particle currents with $\Delta$,  and \textbf{(d)} Variation of max circulating currents with $N_U$. If not  specified otherwise, $N_U=10, N_D=10$. }
		\label{Asymm_figures}
	\end{figure}
	
	where,
	\begin{align}
		v_n=
		\begin{cases}
			t(1-\Delta)& n<N_U+2, \\
			t(1+\Delta)& n \ge N_U+2.
		\end{cases}
	\end{align}
 
	This model can be thought of as connecting two different thermal wires between two heat baths. For electric circuits with this configuration, and following the Ohm's law, we expect the electric current to flow parallelly in the wires. However, this is not the case here as we will see later in this section. The definitions of the currents  for this model  are,
	\begin{align}
		\textnormal{I}_m&=\langle i v_m^2\big{(}\hat{c}_{m+2}^\dagger \hat{c}_m-\hat{c}_{m}^\dagger \hat{c}_{m+2}\big{)}+i \Omega v_m \big{(}\hat{c}_{m+1}^\dagger \hat{c}_m-\hat{c}_{m}^\dagger \hat{c}_{m+1}\big{)}\rangle \nonumber\\
		\textnormal{J}_m&=\langle i v_m \big{(}\hat{c}_{m+1}^\dagger \hat{c}_m-\hat{c}_{m}^\dagger \hat{c}_{m+1}\big{)}\rangle 
	\end{align}

	Using the correlation matrix master equation and performing an analysis similar to that in the SSH model, we obtain the results shown in Fig. \ref{Asymm_figures}.	Looking at the energy spectrum of this model for system size $N=22$  in Fig. \ref{Asymm_figures} (a), we see that there are many AEDPs in this model similar to subsection \ref{subsection:Odd} above. However,  unlike the earlier case all the additional degeneracies do not occur at the same place, and there are many different places where these  degeneracies occur  instead of just the $\Delta=0$. Looking at the corresponding current plots for the setup $N_U=10, N_D=10$ in Fig. \ref{Asymm_figures} (b), we see that CC occurs for this system even for an equal number of Fermionic sites in the branches if the hopping asymmetry parameter $\Delta\ne0$.  Also, the window of $\Delta$ where CC occurs is much wider than in the previous cases. This spread may be accounted to two features of the spectrum, the gradual divergence of degenerate energy levels  from $\Delta=0$ and the presence of extra degeneracies on either side of $\Delta=0$.  
	We also see that as the $\Delta \to 1,-1$, the total current becomes equal to either the upper or lower branch current. This is because in these limits, either the upper or lower branch hopping strength goes to 0.
 
	Observing Fig. \ref{Asymm_figures} (c), we see that the heat and particle current for this system are not exactly equal but the values are very similar.  This tells us that the imaginary part of next to nearest-neighbor correlations are very small in comparison to the imaginary part of nearest-neighbor correlations $Im(\hat{C}_{i,i+2})<<Im(\hat{C}_{i,i+1})$. This also means that most of the energy levels supporting CC are the same for both heat and particle transport.	Finally, we see in Fig. \ref{Asymm_figures} (d) that both the heat and particle CC sustain for larger system sizes in this model. This can be accounted to the increase in number of AEDPs  with system size. However, increasing the number asymmetry in the branches does not necessarily lead to an increase in CC.
	
		\section{Results with interactions turned on} \label{Interaction_section}
		
		\begin{figure}[!htbp]
			\centering 
			\subfigure []
			{\includegraphics[width=0.4\linewidth,height=0.3\linewidth]{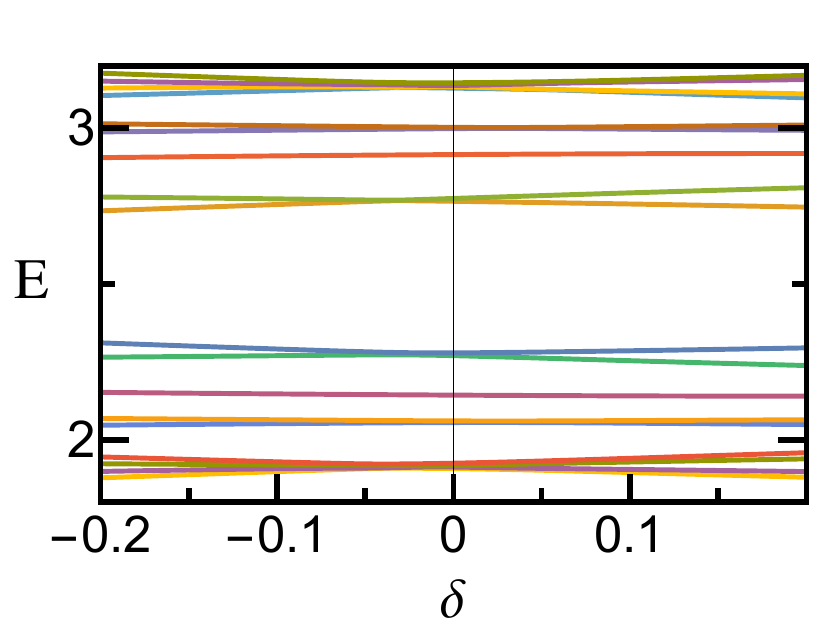}}
			\subfigure []
			{\includegraphics[width=0.4\linewidth,height=0.3\linewidth]{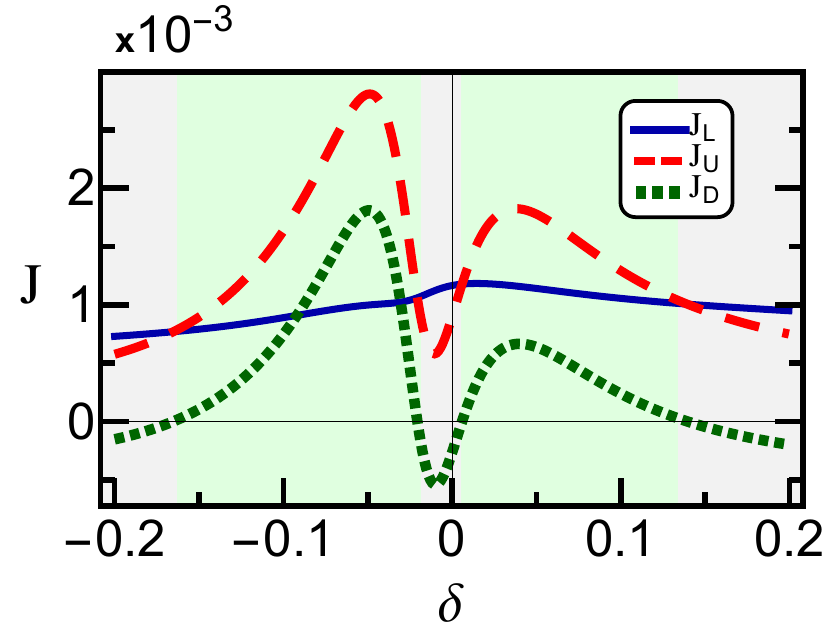}}
			\subfigure []
			{\includegraphics[width=0.4\linewidth,height=0.3\linewidth]{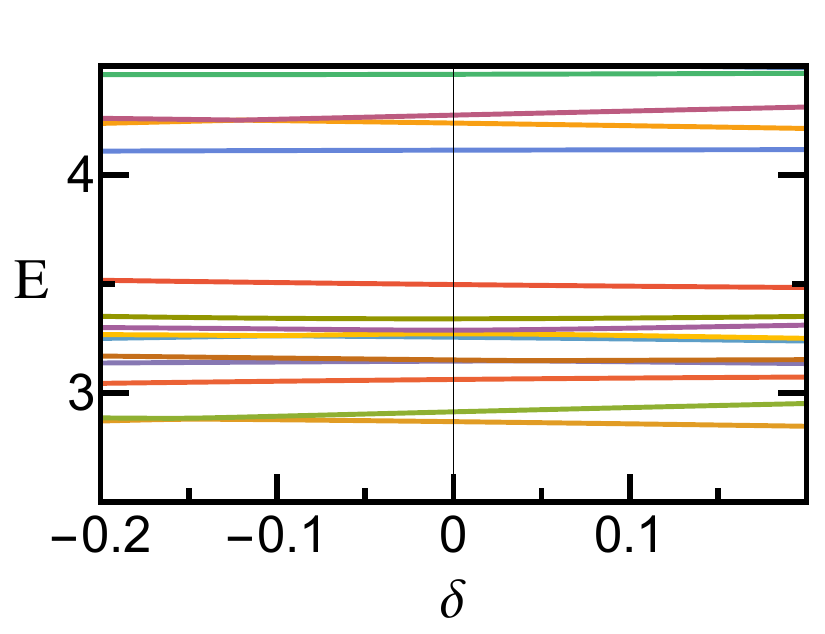}}
			\subfigure []
			{\includegraphics[width=0.4\linewidth,height=0.3\linewidth]{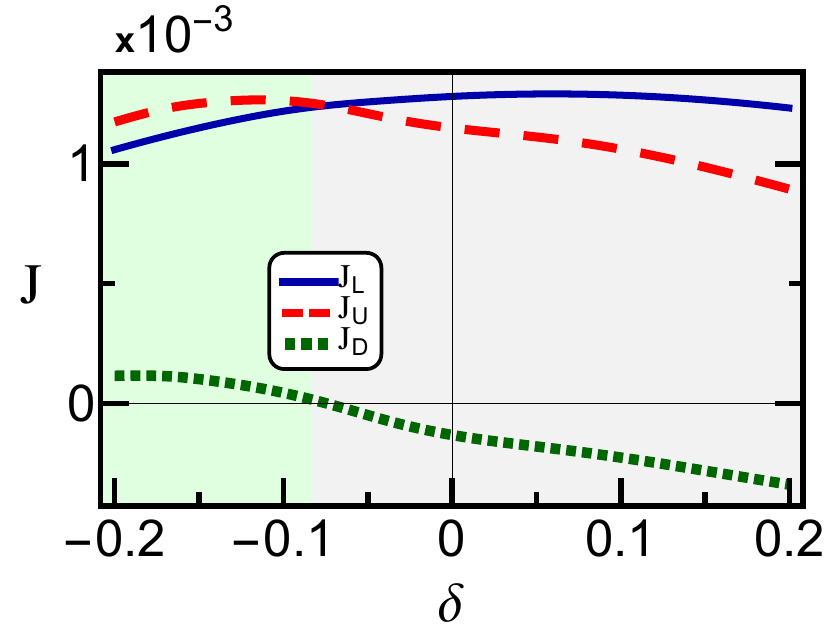}}
			\caption{\textbf{(a)} Variation of the many body energy spectrum with the SSH asymmetry parameter $\delta$ for the modified SSH model containing the additional interaction term given in Eq. \eqref{int_Hamiltonian} for $u=0.02$,  \textbf{(b)} Behavior of the currents with $\delta$ for the weakly interacting modified SSH chain, $u=0.02$. \textbf{(c)} Variation of the many body energy spectrum with the  $\delta$ for the modified SSH model for $u=0.1$,  \textbf{(b)} Behavior of the currents for strongly interacting Fermionic chain with $\delta$ for $u=0.1$.  The values of all other parameters are , $N_U=2, N_D=1, t=0.1, k_{1(4)}=0.01$. For visual convenience we have not plotted the entire many body energy spectrum, and concentrated in a narrow energy channels so that the energy levels are more clearly visible.}
			\label{Interaction_figures}
		\end{figure}
		Finally, we briefly discuss the behavior of the currents when interactions are turned on. To do this we add an interaction term similar to that of the spinless Hubbard model \cite{Hubbard_review_LIEB20031} to the SSH Hamiltonian given in equation \eqref{model_hamiltonian_ssh}. This term has the following form, 
		\begin{align} \label{int_Hamiltonian}
			\hat{H}_I=u\sum_j \hat{P}_j \hat{P}_{j+1} 
		\end{align}
		where, $\hat{P}_j=\hat{c}_j^\dagger \hat{c}_j$ is the particle number operator at site $j$ as defined earlier in eq. \eqref{particle_operator}. Now, since the particle operators at different sites commute, the definition of the particle current in the modified SSH model remains the same as earlier i.e,
		\begin{align}
			\textnormal{J}_m=	\langle \hat{J}_m \rangle =\langle i  t(1+(-1)^m\delta )\big{(}\hat{c}_{m+1}^\dagger \hat{c}_m-\hat{c}_{m}^\dagger \hat{c}_{m+1}\big{)}\rangle
		\end{align}
		Since this is an interacting system, the correlation matrix master equation given in \eqref{Correlation_Matrix_dynamic} is no longer valid. So, we the solve the LLME given in eq. \eqref{local_master_equatio} explicitly with the modified SSH Hamiltonian for the simple small system with $(N_U=2, N_D=1)$. To do this we use the Jordan Wigner transformation \cite{jordan_wigner}, and represent the Fermionic operators, $\hat{c}^\dagger_n (\hat{c}_n)$, in terms of corresponding Pauli matrix representation,
  \begin{align}
      \hat{c}^\dagger_n &= \Pi_{1}^{j<n} \hat{\sigma}^j_z \hat{\sigma}_n^+ \nonumber \\
        \hat{c}_n &= \Pi_{1}^{j<n} \hat{\sigma}^j_z \hat{\sigma}_n^-
  \end{align}
  Here, $\hat{\sigma}^j_z$, is the Pauli-$z$ matrix for the $j^{th}$ spin and $\hat{\sigma}_j^+ (\hat{\sigma}_j^-)$, are the corresponding raising and lowering pauli matrices. The results obtained via this analysis are displayed  in Fig. \ref{Interaction_figures}. We observe in  Fig. \ref{Interaction_figures} (a), that for Fermionic chain where the interaction term is weak in comparison to the hopping strength $(u=0.02,t=0.1)$, the many body energy spectrum still has AEDPs near $\delta=0$, and correspondingly the current circulates near the AEDPs (see Fig. \ref{Interaction_figures} (b)). However, in place of the usual transition from clockwise to anticlockwise current circulation, the transition is clockwise to parallel to again clockwise. 
   Such a transition was also seen in the earlier study \cite{Our_circulation}.  For stronger interaction term in the Hamiltonian $u\sim t$, we see in Fig. \ref{Interaction_figures} (c), that the energy spectrum changes drastically and we no longer have AEDPs near $\delta \sim 0$. Looking at the behavior of currents in Fig. \ref{Interaction_figures} (d), it still seems that the relationship between AEDPs and CC sustains for even strong interactions in the system. However, further analysis is required to theoretically understand the reason behind such a relationship.
		
	\section{Conclusion}\label{Conclusion}
	We study heat and particle CC in quadratic Fermionic systems analysed using a general dissipative Lindblad master equation, with LLME as an example. We solve the correlation matrix master equation perturbatively and obtain an analytical expression indicating the possibility of anomalous  transport like CC near  AEDPs. To investigate this, we study the heat and particle transport in the SSH model and the tight-binding model. We see that just the AEDP is not enough to observe CC and we require some  form of asymmetry in the system. We then apply the geometric asymmetry of having unequal number of Fermionic sites in the upper and lower branches for the SSH model and  observe that CC does get induced near the AEDPs with the direction of CC reversing once we cross the AEDP. {On comparing the results with the exact results obtained via the NEGF formalism, we observe that the relationship between AEDP and CC continues to hold even for the exact results, and like the LLME results, the local branch current changes direction when we cross the AEDPs.} A similar phenomenon is seen in the tight binding model if we employ different hopping strengths in the upper and lower branches. We see that if the number of additional degeneracies increases with system size, the CC remains intact even for larger system sizes, however, if the additional degeneracies do not increase with system size, the CC dies down for larger systems. The width of the parameter region where CC occurs depends on the approach of the otherwise non-degenerate energy levels towards AEDP. If this approach is steep, the window of CC will be narrower and if the approach is gradual, the window of CC will be broader. Finally, we also observe that for certain parameter values, the heat current circulates while the particle current runs parallel in the branches, allowing us to switch on the CC for different types of currents at different points. Adding interactions to the system still seems to sustain the relationship between CC and AEDP, though a more extensive analysis is required to establish this. Considering all of the above analysis, we can conclude that carefully observing  the energy spectrum of the system  gives many indications about the possibility and behavior  of CC in any Fermionic system containing some asymmetry. Though we restrict our study to quadratic Fermionic chains with periodic boundary conditions, this methodology may be useful for studying other forms of anomalous transport in open chains near the AEDPs  with possible  application in sensitive parameter measurements and metrology \cite{Quantum_Wheatstone_Bridge}. 

	\section{Acknowledgement}
	RM gratefully acknowledges financial support from the Science and Engineering Research Board (SERB), India, under the Core Research Grant (Project No. CRG/2020/000620). Authors thank \"Ozg\"ur E. M\"ustecapl\ifmmode \imath \else \i \fi{}o\ifmmode \breve{g}\else \u{g}\fi{}lu for useful discussions and Physics Department, Ko\c{c} University, Istanbul, T\"urkiye for the kind hospitality, where a part of this work was done.
	
	\appendix
	\section{Correlation Matrix Master Equation}\label{Sec:AppendixA}
	The equation governing the matrix element $m,n$ of the correlation matrix is:
	\begin{align}\label{meA}
		\frac{d Tr[\hat{c}_m^\dagger \hat{c}_n \hat{\rho}]}{dt}=&Tr\Big{[}\hat{c}_m^\dagger \hat{c}_n (-i[\hat{H}_S,\hat{\rho}]+\gamma_1 \mathcal{D}(\hat{c}_1)[\hat{\rho}]+ \Gamma_1\mathcal{D}(\hat{c}_1^{\dagger})[\hat{\rho}]+R.B.T\Big{]} 
	\end{align}
	Treating term by term, 
	\begin{align}
		Tr\Big{[}\hat{c}_m^\dagger \hat{c}_n (-i[\hat{H}_S,\hat{\rho}])\Big{]}=-i\sum_{i,j}M_{i,j}\langle [\hat{c}_m^\dagger \hat{c}_n ,\hat{c}_i^\dagger \hat{c}_j ]\rangle 
	\end{align}
	Using the identity, 
	``$	[\hat{c}_m^\dagger \hat{c}_n, \hat{c}_i^\dagger \hat{c}_j]=\delta_{n,i}\hat{c}_m^\dagger \hat{c}_j -\delta_{m,j}\hat{c}_i^\dagger \hat{c}_n$'',
	\begin{align}
		Tr\Big{[}\hat{c}_m^\dagger \hat{c}_n (-i[\hat{H}_S,\hat{\rho}])\Big{]}=-i\sum_j M_{n,j} \langle \hat{c}_m^\dagger \hat{c}_j\rangle -v_{j,m} \langle \hat{c}_j^\dagger \hat{c}_n\rangle
	\end{align}
	We can see that,
	\begin{align}
		Tr\Big{[}\hat{c}_m^\dagger \hat{c}_n (-i[\hat{H}_S,\hat{\rho}])\Big{]}=(i [\hat{M}^T,\hat{C}])_{m,n}
	\end{align}
	where $\hat{M}^T$ is the transpose of the Hamiltonian matrix.
 Now, treating the second term,
	\begin{align}
		Tr\Big{[}\hat{c}_m^\dagger \hat{c}_n \gamma_i \mathcal{D}(\hat{c}_i)[\hat{\rho}]\Big{]}&=-\frac{\gamma_i}{2}(\delta_{i,n}\langle \hat{c}_m^\dagger \hat{c}_i \rangle+\delta_{m,i}\langle \hat{c}_m^\dagger \hat{c}_n \rangle)
		\nonumber \\&=-\frac{1}{2}(\{\hat{R}^i,\hat{C}\})_{m,n}
	\end{align}
	where $\hat{R}^i_{m,n}=\delta_{m,n} \delta_{i,n}\gamma_i$. Hence the combining the contribution from all the dissipators, $\hat{R}=\sum_i \hat{R}^i_{m,n}=\delta_{m,n}\gamma_n$.
	Finally, the last term can be proven to be,
	\begin{align}
		Tr\Big{[}\hat{c}_m^\dagger \hat{c}_n \Gamma_i \mathcal{D}(\hat{c}^\dagger_i)[\hat{\rho}]\Big{]}&=(-\frac{1}{2}\{\hat{G}^i,\hat{C}\}+\hat{G}^i)_{m,n}
	\end{align}
	where, $\hat{G}^i_{m,n}=\delta_{m,n} \delta_{i,n}\Gamma_i$. So, $\hat{G}=\sum_i \hat{G}^i_{m,n}=\delta_{m,n}\Gamma_n$.

	\section {Correlation matrix for quadratic Bosonic systems} \label{Appendix_B}
	Suppose that the operators $\hat{c}_i^\dagger, \hat{c}_i$ in eq. \eqref{meA} are Bosonic and not Fermionic. Also, the Hamiltonian $\hat{H}_S$ is a quadratic Bosonic Hamiltonian. Additionally, the baths are also Bosonic so the functions $\gamma_1$ and $\Gamma_1$ are proportional to the Bosonic distribution functions. Now performing the same analysis as and treating term by term, 
\begin{align}
		Tr\Big{[}\hat{c}_m^\dagger \hat{c}_n (-i[\hat{H}_S,\hat{\rho}])\Big{]}=-i\sum_{i,j}M_{i,j}\langle [\hat{c}_m^\dagger \hat{c}_n ,\hat{c}_i^\dagger \hat{c}_j ]\rangle 
	\end{align}
	Here, using the Bosonic commutation identity, 
	$	[ \hat{c}_i, \hat{c}^\dagger_j]= \delta_{i,j}$, we can see that,
	\begin{align}
		Tr\Big{[}\hat{c}_m^\dagger \hat{c}_n (-i[\hat{H}_S,\hat{\rho}])\Big{]}=(i [\hat{M}^T,\hat{C}])_{m,n}
	\end{align}
	where $\hat{M}^T$ is the transpose of the Hamiltonian matrix.
	Now, treating the second term, we get the same expression contribution,
	\begin{align}
		Tr\Big{[}\hat{c}_m^\dagger \hat{c}_n \gamma_i \mathcal{D}(\hat{c}_i)[\hat{\rho}]\Big{]}&=-\frac{\gamma_i}{2}(\delta_{i,n}\langle \hat{c}_m^\dagger \hat{c}_i \rangle+\delta_{m,i}\langle \hat{c}_m^\dagger \hat{c}_n \rangle)
		\nonumber \\&=-\frac{1}{2}(\{\hat{R}^i,\hat{C}\})_{m,n}
	\end{align}
	where $\hat{R}^i_{m,n}=\delta_{m,n} \delta_{i,n}  \gamma_i$.
	\bibliography{circulation}
	
\end{document}